\documentclass{article}

\usepackage{PRIMEarxiv}

\usepackage[utf8]{inputenc} 
\usepackage[T1]{fontenc}    
\usepackage{hyperref}       
\usepackage{url}            
\usepackage{booktabs}       
\usepackage{amsfonts}       
\usepackage{nicefrac}       
\usepackage{microtype}      
\usepackage{lipsum}
\usepackage{fancyhdr}       
\usepackage{graphicx}       
\graphicspath{{media/}}     

\usepackage[longnamesfirst,sort]{natbib}
\bibpunct[, ]{(}{)}{;}{a}{,}{,}
\renewcommand\bibfont{\fontsize{10}{12}\selectfont}

\usepackage[normalem]{ulem}
\useunder{\uline}{\ul}{}

\usepackage[utf8]{inputenc} 
\usepackage[T1]{fontenc}    
\usepackage{hyperref}       
\usepackage{url}            
\usepackage{booktabs}  
\usepackage[table,xcdraw]{xcolor}
\usepackage{xcolor}
\usepackage{amsfonts}       
\usepackage{nicefrac}  
\usepackage{mathrsfs}
\usepackage{amsmath}
\DeclareMathOperator*{\argmax}{argmax}
\usepackage{graphicx}
\usepackage{commath}
\usepackage{amssymb}
\usepackage{multirow}
\usepackage{microtype} 
\usepackage{bbm}
\usepackage{lipsum}
\usepackage{tikz}
\usepackage{xcolor}
\usetikzlibrary{automata, arrows,shapes,positioning,shadows,trees}

\usetikzlibrary{shapes,snakes}
\usetikzlibrary{shapes,arrows}
\usetikzlibrary{calc,trees,positioning,arrows,chains,shapes.geometric,%
    decorations.pathreplacing,decorations.pathmorphing,shapes,%
    matrix,shapes.symbols}

\newcommand{\blue}[1]{{\color{black} #1}}

\begin{document}




\title{Efficient global estimation of conditional-value-at-risk through stochastic kriging and extreme value theory
}

\author{
  Armin Khayyer \\
  Department of Industrial and Systems Engineering \\
  Auburn University \\
  Auburn\\
   \And
  Alexander Vinel \\
  Department of Industrial and Systems Engineering \\
  Auburn University \\
  Auburn\\
  \texttt{alexander.vinel@auburn.edu} \\
     \And
  Joseph J. Kennedy \\
  Department of Industrial and Systems Engineering \\
  Auburn University \\
  Auburn\\
}

\maketitle

\begin{abstract}
We consider the problem of evaluating risk for a system that is modeled by a complex stochastic simulation with many possible input parameter values. Two sources of computational burden can be identified: the effort associated with extensive simulation runs required to accurately represent the tail of the loss distribution for each set of parameter values, and the computational cost of evaluating multiple candidate parameter values.  The former concern can be addressed by using Extreme Value Theory (EVT) estimations, which specifically concentrate on the tails. Meta-modeling approaches are often used to tackle the latter concern. In this paper, we propose a framework for constructing a particular meta-modeling framework, stochastic kriging, that is based on EVT-based estimation for a class of coherent measures of risk. The proposed approach requires an efficient estimator of the intrinsic variance, and so we derive an EVT-based expression for it. It then allows us to avoid multiple replications of the risk measure in each design point, which was required in similar previously proposed approaches, resulting in a substantial reduction in computational effort. We then perform a case study, outlining promising use cases, and conditions when the EVT-based approach outperforms simpler empirical estimators.\end{abstract}

\keywords{risk measures \and Conditional Value-at-Risk \and stochastic kriging \and metamodelling \and simulation}

\section{Introduction}
It is often the case that in modeling a stochastic system the decision-maker has no prior knowledge of the response distribution, and instead, through observing the outputs in a simulation or through sampling, they can characterize and quantify its tail behavior, and consequently risk. In many disciplines, minimizing the impact of rare catastrophic losses exceeding a certain threshold is of great importance, and consequently, risk measures designed for quantifying the tail behavior are well-studied in the risk management literature. \blue{Conditional Value-at-Risk ($\mathrm{CVaR}$) (which is the main focus of this effort), for instance,  a coherent risk measure introduced by \cite{rockafellar2000optimization}, has gained particular attention as it has been  shown to possess significant advantages over other approaches, most notably, Value-at-Risk \citep{krokhmal2013modeling}. }

It is well known that parametric or non-parametric estimation of the  entire density is ill-suited for estimating extreme quantiles of losses. Such  methods tend to  find a good fit where most of the data falls and ignore the details of the tail behavior. Extreme value theory (EVT), on the other hand, focuses on the extreme outcomes specifically and tries to approximate only the tail distribution, instead of the entire region. This approximation can be valid for any value larger than a certain threshold, and can sometimes be used for  regions even outside the range of the observations. EVT is widely used in  varied fields, for example, finance or hydrology.  While it is also known to improve tail estimates for certain stochastic systems \citep{mcneil1997peaks}, it has not, so far, attracted a sufficient  amount of attention in the  simulation and stochastic optimization communities. 
The Peak Over Threshold (POT) is  a particular method to model extreme
values, which aims to describe the asymptotic tail distribution of a random variable for values above a threshold. It can be demonstrated that the
distribution of the values exceeding the threshold can be approximated by a General Pareto
Distribution ($\mathrm{GPD}$) for some sufficiently large threshold  \citep{balkema1974residual,pickands1975statistical}. 

In practice, when assessing the risk measures and uncertainty of a stochastic simulation model is of interest, POT is implemented by drawing a large enough sample of the simulation output to ensure that a sufficient number of rare events are observed. Then, the tail behavior is approximated using this sample, and consequently, the risk measures are computed. This is even more computationally expensive when the simulation model covers  a  wide  range of parameters. Moreover, conducting a thorough analysis of the risk measure for every combination of input parameters of the model becomes quickly prohibitive as assessing the risk measure for each requires many function evaluations and replications. This has resulted in significant attention paid to the  area  of meta-modeling, or surrogate modeling,  see \citet{barton2006metamodel} for a thorough review.




\blue{In this work, we present a  metamodeling framework for efficient global prediction of risk across the problem domain. We specifically focus on a particular popular tail measure, Conditional Value-at-Risk ($\mathrm{CVaR}$), and derive the necessary expressions, before expanding it to a more general case of all law-invariant comonotone additive measures of risk. The approach is based on employing a combination of Stochastic kriging and EVT-based estimation of tail behavior, building on the results first discussed in \cite{9384101} a simple application of EVT as a building block for $\mathrm{CVaR}$ estimation in a kriging model was presented. 

We expand on the framework in three important ways. First, we present a comprehensive discussion of how EVT methodology can be used for tail estimation, which includes analytical formulae for estimating the variance of $\mathrm{CVaR}$ at each design point using POT and the asymptotic properties of the Maximum Likelihood Estimation of the $\mathrm{GPD}$ parameters. The estimated variance of the risk measure is subsequently employed within the context of Stochastic kriging as intrinsic noise. This enables the kriging model to more efficiently use the simulation budget without the need for replicating $\mathrm{CVaR}$ estimates at each design point to extract an empirical variance estimate, compared to the previously reported results. Second, we expand the definition to the  more general case of law-invariant and comonotone additive coherent measures of risk. Finally, we perform more a thorough testing of the approach, comparing it against both the previous version, standard ordinary kriging as well as another specialized approach from the literature. }

\subsection{Brief Review of Related Literature}

 Building a metamodel for quantile-based risk measures is not a new research avenue. \cite{koenker2001quantile} proposed quantile regression which estimates the conditional median (or other quantiles) of the response variable instead of the conditional mean. In the same direction, \cite{dabo2010robust} proposed a kernel-based statistical estimator for modeling conditional quantiles of spatial processes. \cite{liu2009estimating} proposed a two-level procedure to estimate the expected shortfall (ES) also known as $\mathrm{CVaR}$ (note that in this work, we prefer to use the term $\mathrm{CVaR}$, which sometimes is used interchangeably with ES and AVaR,  Average Value-at-Risk). Their procedure first constructs a meta-model based on an initial set of design points and then uses the posterior distribution to learn the scenarios that are most likely to
entail large losses. It then adds the likeliest points to the design point set and allocates the remaining computational budget accordingly. Another metamodel is constructed at the end of the second stage and is used to estimate  $\mathrm{CVaR}$. \cite{begueria2006mapping} proposed a procedure based on extreme value analysis and spatial interpolation techniques. The parameters of $\mathrm{GPD}$ were regressed over the inputs, resulting in a probability model in which the distribution parameters vary smoothly in space. 

\cite{reza2013analysis} developed a spatial model of the parameters of an extreme distribution (i.e. $\mathrm{GPD}$) through a hierarchical Bayesian process with latent parameters considered as random variables and simulated using Markov Chain Monte Carlo techniques. Various prior distributions are assumed for the $\mathrm{GPD}$ parameters and used to assess the sensitivity of the posterior distributions. \cite{bracken2018bayesian} modeled multivariate nonstationary frequency for hydrologic variables using a Bayesian hierarchical framework. The joint distribution of all variables is modeled by a Gaussian elliptical copula, and it is assumed that the marginal distribution of each variable follows a generalized extreme value (GEV) distribution, where the GEV parameters can vary in time. \cite{wang2017joint} extended the standard stochastic Gaussian process model to the multi-response case to model jointly the quantile (estimated using sectioning) with a correlated and less-noisy expectation to improve the fit and predictions of the quantile metamodel.

\cite{chen2016efficient} extended the stochastic kriging (SK) model to approximate quantiles from stochastic simulation models. The authors analyzed the impact of different non-parametric 
point and variance estimators of VaR and $\mathrm{CVaR}$ such as batching \citep{rougier2008efficient}, sectioning \citep{asmussen2007stochastic}, sectioning-batching \citep{mitchell1994asymptotically}, and jackknifing \citep{kim2007quantifying} on the predictive performance of SK. Furthermore, SK has been shown to be a powerful technique that extends the applicability of deterministic kriging metamodeling to
modeling responses from stochastic simulations, including forecasting tail measures \citep{chen2016efficient}. \cite{9384101} explicitly concentrated on estimating $\mathrm{CVaR}$ as the primary measure of risk and tail behavior. POT method was used to estimate $\mathrm{CVaR}$ at each design point, while the variance of $\mathrm{CVaR}$ was estimated empirically by replicating the POT $\mathrm{CVaR}$ estimation.  

\blue{
We discuss specific differences in some of the existing POT- and empirical-based estimation schemes in more detail in Section \ref{method}, when presenting the proposed approach.  Most importantly, we demonstrate that
parametric estimation 
can be simultaneously accomplished for both risk value and its variance. Specifically, instead of replicating the estimation of $\mathrm{CVaR}$ at each design point to estimate
its variance, we exploit the asymptotic behavior of the maximum likelihood estimators of the parameters of the $\mathrm{GPD}$ to construct an estimate for the variance of  $\mathrm{CVaR}$. This then means that a single replication can be sufficient. 
Consequently, we contribute to the literature on efficiently estimating $\mathrm{CVaR}$ by employing a two-stage framework, coupling metamodeling with the parametric estimation of both $\mathrm{CVaR}$ and its variance using EVT which distinguishes our effort from the existing literature and more specifically from \citep{chen2016efficient, 9384101}. 
Further, we show that in principle any coherent, law-invariant, and comonotonic additive risk measure is amenable to a similar approach, by analyzing the spectral representation of such measures. Note, though, that for any given measure it may be possible to construct a more efficient and straightforward estimator, compared to the generic approach considered here.
}

\subsection{Organization}
The remainder of this paper is organized as follows. In Section \ref{sec:cvar-risk-measures}, we present an overview of EVT and stochastic kriging, specifically as applied to $\mathrm{CVaR}$. We present the proposed method of estimating $\mathrm{CVaR}$ and its variance using EVT in Section \ref{method}. Lastly, in Section \ref{experandres}, we present two numerical examples, their experimental designs, and results. We compare the results of our proposed method with that of \cite{9384101} and  \cite{chen2016efficient}, and we show promising results with the parametric estimation of $\mathrm{CVaR}$ and its variance. 

\section{Background}
\label{sec:cvar-risk-measures}
\subsection{$\mathrm{CVaR}$ and its estimation}
\label{sec:cvar}
$\mathrm{CVaR}$ is a risk assessment measure that quantifies the amount of tail risk. For a given random variable $X$, with its cumulative distribution function (CDF) given by $F_X$ and its probability density function (PDF) given by $f_X$, $\mathrm{CVaR}$ at level  $\alpha$ is denoted as $CVaR_\alpha (X)$. Recall that the $\alpha$-quantile  of X is also known as Value at Risk (VaR): 
$q_\alpha = inf\{x\in R: F_X(x)\geq \alpha\}$.
If the random variable $X$ is absolutely continuous,  $\mathrm{CVaR}$  can be defined as the expectation of values exceeding value-at-risk: 
\begin{equation}
    CVaR_\alpha (X)=E\Big[X|X>q_\alpha \Big] = \dfrac{1}{1-\alpha}\int_{q_\alpha}^{+\infty}{xf_X(x)dx}.
    \label{eq:cvarintegral}
\end{equation}
For details on the definition in the case of a non-continuous distribution, consider \citet{rockafellar2002conditional}. 

Since in practice the CDF, $F_X$, is often unknown, an empirical distribution function can be used to describe a sample of observations of variable $X$.  Consider an i.i.d. sample  $ S_t = \{x_1, . . . , x_t\}$ of observations drawn from distribution
$F_X$. The empirical distribution function is defined as: $\widehat{F}_X^t = t^{-1} \sum_{n=1}^t \mathbbm{1} _{\{x_n \leq x\}}$.
\blue{Thus the quantile value, $q_\alpha$, at level $\alpha$ can be estimated using the empirical CDF as

$$\widehat{q}_\alpha = inf\{u: \widehat{F}(u) \geq \alpha \}$$

leading to the empirical $CVaR_\alpha(X)$ defined as 
\begin{equation}
    \widehat{CVaR_\alpha}(X)= \frac{\sum^N_{i=1} x_i \mathbbm{1}_{\{ x_i\geq \widehat{q}_\alpha\}}}{\sum^N_{i=1} \mathbbm{1}_{\{ x_i\geq \widehat{q}_\alpha\}}}.
    \label{eq:emp}
\end{equation}
}

Typical values of $\alpha$ of interest are  usually small (e.g. less than 5\%), meaning that empirical CDF, in this case, can be inaccurate. Indeed, small $\alpha$ leads to few exceedances, and hence, small  sample size (and so high variance) for the estimator for VaR and $\mathrm{CVaR}$.
 Extreme Value Theory (EVT) attempts to approximate $\mathrm{CVaR}$ by fitting a model to only the tail of the distribution from scarce samples by exploiting the asymptotic behavior of the tail distribution above a high
threshold, $u$, and then extrapolating the conditional expectation directly from it. The corresponding exceedance distribution is given by

\begin{equation}
\begin{aligned}
F_u (z) &=  \mathbb{P}(X-u\leq z |X > u)
 = \dfrac{\mathbb{P}(X-u\leq z , X > u)}{\mathbb{P}(X > u)} \\
& = \dfrac{\mathbb{P}(u<X\leq z + u)}{\mathbb{P}(X > u)}
= \frac{F(z+u)-F(u)}{1- F(u)},
\end{aligned}
\label{eq:FU}
\end{equation}
where the threshold excesses are denoted with $z$ value and $F_u$ shows the conditional probability that $X$ exceeds the threshold, $u$, by at most $z$. While  the distribution function is often unknown, the exceedance distribution can be shown (under some conditions) to be well-described with Generalized Pareto Distribution ($\mathrm{GPD}$) with appropriately selected parameters. 
$\mathrm{GPD}$ with two parameters, shape  $\xi$ and scale  $\beta > 0$, is a family of continuous probability functions defined by the following cumulative distribution function:
\begin{equation}
G_{\xi, \beta}(x) = 
  \begin{cases}
       $$1- (1+\frac{\xi x}{\beta})^{\frac{-1}{\xi}}$$ & $$\text{if }  \xi \neq 0$$\\
      $$ 1- e^{\frac{-x}{\beta}}$$ & $$\text{if }  \xi = 0$$.\\
      \end{cases}
\end{equation}
The following fundamental result characterizes the distribution of exceedances as a $\mathrm{GPD}$.

\noindent\textbf{Theorem 1    \citep{balkema1974residual,pickands1975statistical}} Consider a real value $\xi$ and a random variable $Y$ and let $x_0$ be the right endpoint (finite or infinite) of distribution $F(x)$ and $0\leq u< x_0$. Then a positive measurable function $\beta(u)$ can be found such that, 
\begin{equation}
 \lim_{u \rightarrow x_0} \sup_{0 \leq x\leq x_0-u} |F_u(x)-G_{\xi,\beta(u) }(x) |= 0,
\end{equation}
if and only if there exist real sequences $a_n$ and $b_n$ such that $F^n (a_n x+b_n )\rightarrow H_\xi (x)$ as $n\rightarrow \infty$, for all $x \in \mathbb{R}$. Where $F^n(x) = \mathcal{P}(\max(X_1, . . . , X_n) \leq x)$ denote the cdf of the sample maxima.  Note that $F$ is the CDF of some random variable and $H_\xi$ is the Generalized Extreme Value (GEV) CDF with parameter $\xi$. Then F belongs to the Maximum Domain of Attraction (MDA) of $H_\xi$.



In other words, Theorem 1 states that the distribution of
the values exceeding a large enough threshold $u$ can be approximated by a
$\mathrm{GPD}$ with shape $\xi$ and scale
$\beta$ parameters. 
\blue{Note that the inverse of \eqref{eq:FU} gives the high quantile of the
distribution or $VaR$. For $\alpha \geq F(u)$,  $VaR$ is given by

\begin{equation}
    VaR_{\alpha} = u + \frac{\beta}{\xi} \bigg[ 
\Big(\frac{1-\alpha}{F(u)} \Big)^{-\xi} - 1
\bigg]
\label{eq:VaR}
\end{equation}
}

Consequently, using equation \eqref{eq:cvarintegral}, we can estimate the conditional value at risk as: 
\begin{align*}
CVaR_\alpha (X) = \widehat{q}_\alpha + \frac{\widehat{\beta}(u)- \widehat{\xi}(\widehat{q}_\alpha -u)}{1-\widehat{\xi}},
    \label{eq:cvar}
\end{align*}
where $q_\alpha$ is the quantile of confidence
level $\alpha$ of $X$, $u$ is a sufficiently large  threshold such that $u \leq q_\alpha$, and $\widehat{\beta}(u)$ is the function used in Theorem 1. 
This estimation approach is  known as Peak Over Threshold-method (POT method) which is one way to model the extreme
values. For more details, refer to \cite{mcneil1997estimating}. 

\cite{de2006extreme} demonstrated that the exceedances over a given threshold $u$ are i.i.d and approximately distributed by a $\mathrm{GPD}$. They also showed that the $\mathrm{GPD}$ parameters can be estimated using the maximum likelihood estimator (MLE) and proved its asymptotic normality. \cite{smith1985maximum} showed that when $\xi < \frac{1}{2}$, the usual
properties of consistency, asymptotic normality, and
asymptotic efficiency hold for the estimators. MLE estimates can be obtained by solving the following optimization problem
\begin{equation}
    (\widehat{\xi}, \widehat{\beta}) = \argmax_{\xi, \beta} \sum_{z \in Z_u}\log \ g_{\xi, \beta}(z),
\end{equation}
where $g_{\xi, \beta}(z)$ is the PDF of the continuous generalized Pareto distribution ($\mathrm{GPD}$) with two parameters $\xi \in \mathbb{R}$ and
$\beta >0$. Using computer simulation, \cite{hosking1987parameter} showed that the MLE estimator can exhibit good performance, and such approaches have been used instead for empirical methods in many applications. Note that since the MLE optimization problem does not have a closed-form analytical solution, it must be calculated numerically. \blue{It is worth mentioning that, in the context of $\mathrm{CVaR}$ estimation with EVT, as long as a suitable threshold is chosen, the MLE estimators are asymptotically normal, but with a biased mean \citep{de2006extreme}. Note that \citet{troop2021bias} provided a bias correction approach for estimating $\mathrm{GPD}$ parameters. As a result, it can be  additionally shown that a much lower threshold can be chosen
without presenting significant bias if such a corrected approach is employed. }

\subsection{Stochastic kriging}

The basic theory of kriging for deterministic modeling was extended to the stochastic modeling setting by \cite{4736089}. To accomplish this, both intrinsic and extrinsic uncertainties are considered, where the former characterizes the inherent stochasticity of the simulation model, and the latter is due to the unknown response surface. Let  $X=[x^{(1)},x^{(2)},...,x^{(k)}]^T$ be a training set, where $x^{(i)}=[x_1^{(i)}, x_2^{(i)},...,x_d^{(i)}]$, $i=1,2,...,k$ are $d$-dimensional design points, which can be specified using an experiment design. Note that as we are interested in stochastic modeling, $\Bar{Y}^{(i)}$ denotes the average of observations over the entire number of replications at each design point, and the observed response vector for all of the $k$ design points is $ \Bar{Y} =[\Bar{Y}^{(1)}, \Bar{Y}^{(2)}, ..., \Bar{Y}^{(k)}]^T$.  

Given the notation above, the response value of a stochastic simulation model of replication $j$ at design point $x$ can be represented as:
\begin{equation}
    y_j (x)=f(x)^T \beta+M(x)+ \epsilon_j (x),
\end{equation}
where $M(x)$, a mean zero stationary Gaussian random field (GRF), is referred to as extrinsic uncertainty that represents fluctuations around the trend and exhibits spatial correlation. That is, if two points are spatially located close to each other their value of $M(x)$ will be similar.
$M$, defined as a zero mean Gaussian random field, implies that $M(x^{(1)}), M(x^{(2)}), \dotsc, M(x^{(k)})$ are multivariate normal for every collection of design points from a discrete domain $(x^{(1)}, x^{(2)}, \dotsc, x^{(k)})$. Therefore, $M$ can be completely characterized by its covariance matrix. Note that the covariance of the two random variables, $M(x^i) \text{ and } M (x^j)$, only depends on the distance between $x^i$ and $x^j$.  $\epsilon(x)$ is intrinsic noise. It is assumed that the variance of noise at each design point is not constant and $Corr[\epsilon(x), \epsilon(x')] >0$. Moreover, $\epsilon_1(x^i), \epsilon_2(x^i), \dotsc \epsilon_{n^i}(x^i)$ are i.i.d. with $N(0, V(x^i))$ where $n_i$ shows the number of replications at that design point. Lastly, $f(x)^T$ is a $(m \times 1)$ vector of polynomial basis functions.

For a collection of design points $x^1, x^2, \dotsc x^k$, let $\Sigma_{M}(x^i,x^j) = Cov[M(x^i),M(x^j)]$ be a $k\times k$ \blue{covariance matrix which is obtained by the extrinsic spatial correlation function $r(x^i,x^j, \theta)$.} Moreover, let $x_0$ be a new location in the domain and $\sum_M(x_0, \cdot )$ be a $k\times1$ vector
$[Cov[M(x_0),M(x^1)], Cov[M(x_0),M(x^2)], \dotsc, Cov[M(x_0),M(x^k)]]$. 
 Let  $\Sigma_\epsilon$ be the $k\times k$ intrinsic covariance matrix. Assuming that $\beta, \Sigma_M, \Sigma_\epsilon$ are known \citep{4736089}, the SK predictor is then given as follows: 
  \begin{equation}
    \widehat{y}(x_0) = f(x_0)^T \beta+ \Sigma_M (x_0,.)^T  [\Sigma_M+ \Sigma_\epsilon ]^{-1} (\Bar{Y}-F\beta),
    \label{eq:pred}
\end{equation}
where $F$ is a matrix with its $i^{th}$ row given by $F_i=f(x_i)$. Assume that $M(x)$ is second-order stationary, that is, the expectation and variance of $M(x)$, respectively $E[M(x)]$ and $Var[M(x)]$, are constant over the entire study domain, or in other words, they do not depend on the location $x$ inside the field. Then $\Sigma_{M}(x,x')$ can be rewritten as $\tau^2 r_M (x,x':\theta)$. Therefore as mentioned earlier, the covariance between two observations separated by a distance $h$, $cov(M(x+h),M(x))$ only depends on the distance $h$ between the locations and not on the spatial location $x$ inside the field.
Now, let $R_M(\theta)$ be $\Sigma_M/ (\tau^2)$ then equation \eqref{eq:pred} can be reformatted as:  
\begin{equation}
    \widehat{y}(x_0) = f(x_0)^T \beta+ r(x_0, .)\Big[R_M(\theta)+ \frac{\Sigma_\epsilon}{\tau^2}\Big]^{-1} (\Bar{Y}-F\beta).
\end{equation}

Note that, when $\Sigma_\epsilon$ (the intrinsic covariance matrix) is negligible compared to the extrinsic variance, the above equation reduces to the equation of Universal kriging and the model will predict a weighted average of the $\Bar{Y}$ for any given location $X_0$ which is not in the set of design points. Moreover, if $\tau^2$ (also known as extrinsic variance) is negligible and close to zero, then the above equation approaches the polynomial regression $\widehat{y}(x)=f(x)^T \beta$, where $f(x)$ shows the different features of the deterministic regression model.
As stated earlier, it is assumed that  $\epsilon_1(x_i), \epsilon_2(x_i), \dotsc, \epsilon_n(x_i)$ are i.i.d and normally distributed $\mathcal{N}(0, V(x_i))$. This assumption can often be justified since the output of a complex simulation model is itself the average of many basic random variables. It is demonstrated that $V(x_i)$ can be estimated using the following equation,
\begin{equation}
    \widehat{V}(x_i) = \dfrac{1}{n_i - 1}\sum_{j=1}^{n_i} (y_j(x_i) - \Bar{Y}(x_i))^2.
    \label{eq:var-emp}
\end{equation}

Assuming $\Sigma_\epsilon$ is known and equal to $Diag\{V_1,V_2,...,V_k\}$ where $V_i=  \widehat{V}(x_i)/n_i$, $n_i$ gives the number of replications at design point $x_i$, and considering the normality assumptions of the kriging predictor, the log-likelihood function of the collected data is  \citep{4736089}: 
\begin{equation}
	\ell(\beta,\theta,\tau^2)= -\ln[(2\pi)^{\frac{k}{2}}] -\frac{1}{2} \ln[|\tau^2 R_M (\theta) + \Sigma_\epsilon|] -  1/2 (\Bar{Y} - F\beta)^T [\tau^2 R_M (\theta)+ \Sigma_\epsilon ]^{-1} (\Bar{Y}-F\beta ).
	\label{eq:likelihood}
\end{equation}

Since $R$ may be nearly singular and $\tau^2$ may be small, maximization over the three parameters simultaneously is unstable \citep{fang2005design}. It is demonstrated in practice that the maximum likelihood estimator of $\beta$ is asymptotically independent of $(\tau^2, \theta)$ \blue{\citep[refer to][for more details]{fang2005design}}. Therefore, one can separately and iteratively estimate $\beta$ and $(\tau^2, \theta)$. Note that unlike $\beta $ and $\tau^2$, $\theta$ does not have a closed-form estimator and it requires a numerical search algorithm to maximize the likelihood with respect to $\theta$. It is worth mentioning that since $\Sigma_\epsilon$ is not a function of the parameters, adding the diagonal intrinsic noise matrix, $\Sigma_\epsilon$, to the extrinsic covariance matrix, $\Sigma_M$, works in our favor, as it makes $\Sigma = \Sigma_M + \Sigma_\epsilon$ resistant to becoming singular. This may allow maximizing the likelihood function simultaneously over the parameters of the kernel, $\theta$, the extrinsic variance of the GRF, $\tau^2$, and finally the coefficients of the deterministic trend,  $\beta$.

\section{Methodology}\label{method}
\subsection{Overall framework}
In a risk-driven stochastic simulation system, the decision-makers are interested in describing tail behavior for the underlying distributions of the outputs of the simulation model. Depending on the distribution, which is unknown in most cases, characterizing risk requires many function evaluations at each combination of the input parameters of the system. The idea in this work is to construct a metamodel that can be used globally across the domain of the simulation model, allowing estimation of the risk measure (i.e. $\mathrm{CVaR}$) for any given input in real-time. This task is otherwise computationally expensive if the simulation model itself is used to estimate risk measures for any input, as it requires many observations at each point.

\tikzstyle{decision} = [diamond, draw,fill=black!20, 
    text width=7em, text badly centered, node distance=3cm, inner sep=0pt]
\tikzstyle{block} = [rectangle, draw, fill=blue!20, 
    text width=26em, text centered, rounded corners, minimum height=3em]
\tikzstyle{line} = [draw, -latex']
\tikzstyle{cloud} = [draw, ellipse,fill=red!20, node distance=3cm,
    minimum height=2em]
    
\tikzset{pblock/.style = {rectangle split, rectangle split horizontal,
                      rectangle split parts=2, very thick,draw=black!50, top
                      color=white,bottom color=black!20, align=center}}
\tikzset{
>=stealth',
  punktchain/.style={
    rectangle, 
    rounded corners, 
    draw=black, very thick,
    text width=10em, 
    minimum height=3em, 
    text centered, 
    on chain},
  line/.style={draw, thick, ->},
  element/.style={
    tape,
    top color=white,
    bottom color=blue!50!black!60!,
    minimum width=8em,
    draw=blue!40!black!90, very thick,
    text width=10em, 
    minimum height=3.5em, 
    text centered, 
    on chain},
  every join/.style={->, thick,shorten >=1pt},
  decoration={brace},
  tuborg/.style={decorate},
  tubnode/.style={midway, right=10pt, below =10pt},
}
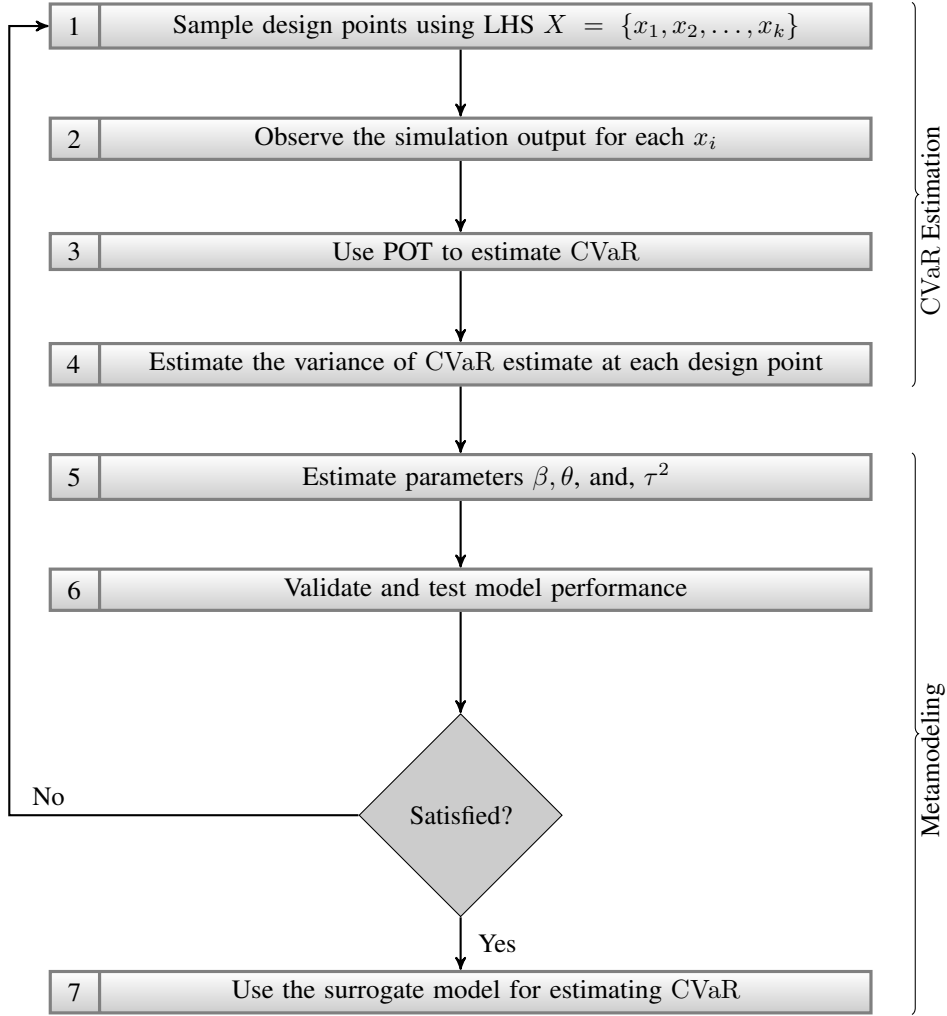
\begin{figure}[!h]
    \centering
\begin{tikzpicture}[node distance = 1.5cm, auto]

     \node[pblock](init){\nodepart[text width=.4cm]{one} 1
                \nodepart[text width = 10cm]{two}Sample design points using LHS
    $X = \{x_1, x_2, \dots , x_k\}$};
    \node[pblock, below of=init](identify){\nodepart[text width=.4cm]{one} 2
                \nodepart[text width = 10cm]{two}Observe the simulation output for each $x_i$};
    
    \node[pblock, below of=identify](POT){\nodepart[text width=.4cm]{one} 3
                \nodepart[text width = 10cm]{two}Use POT to estimate $\mathrm{CVaR}$};
                
    \node[pblock, below of=POT](estimate){\nodepart[text width=.4cm]{one} 4
                \nodepart[text width = 10cm]{two}Estimate the variance of $\mathrm{CVaR}$ estimate at each design point};
    
    \node[pblock, below of=estimate](parameters){\nodepart[text width=.4cm]{one} 5
                \nodepart[text width = 10cm]{two}Estimate parameters $\beta, \theta$, and, $\tau^2$};
    
    \node[pblock, below of=parameters](validate){\nodepart[text width=.4cm]{one} 6
                \nodepart[text width = 10cm]{two}Validate and test model performance};
                
    \node[decision, below of= validate](Satisfied){Satisfied?};
    \node[pblock, below=.7cm of Satisfied](surrogate){\nodepart[text width=.4cm]{one} 7
                \nodepart[text width = 10cm]{two}Use the surrogate model for estimating $\mathrm{CVaR}$};

    \path [line] (init) -> (identify);
    \draw[tuborg, decoration={brace}] let \p1=(init.north), \p2=(estimate.south) in
    ($(6, \y1)$) -- ($(6, \y2)$) node[tubnode, rotate=90] {$\mathrm{CVaR}$ Estimation};
    \path [line] (identify) -- (POT);
    \path [line]  (POT) -- (estimate);
    \path [line]  (estimate) -- (parameters);
    \path [line]  (parameters) -- (validate);
    \path [line]  (validate) -- (Satisfied);
    \draw[tuborg, decoration={brace}] let \p1=(parameters.north), \p2=(surrogate.south) in
    ($(6, \y1)$) -- ($(6, \y2)$) node[tubnode, rotate=90] {Metamodeling};
   \path[line] (Satisfied) -> (surrogate) node [tubnode, rotate=0, right= .1cm] () {Yes};
    \path[line] (Satisfied.west) -- (-6, -10.5) -- (-6, 0) -> (init.west) node[ below=10cm] () {No}; 
\end{tikzpicture}
 \caption{Overall metamodeling framework}
\label{fig:methodology}
\end{figure}


The overall metamodeling framework (Figure \ref{fig:methodology}) consists of two major components, namely risk estimation and construction of a global predictor. The first step in the risk estimation phase is to sample a collection of design points. For each point in this set, the simulation response is observed multiple times to enable estimation of the risk measure, $\mathrm{CVaR}$, using POT. Note that in addition to the POT estimate,  $\mathrm{CVaR}$ value can also be estimated empirically, i.e. as the average of exceedances, equation \eqref{eq:emp}, similar to \citet{chen2016efficient}. The estimate of $\mathrm{CVaR}$ is highly dependent on the underlying distribution and the samples. Therefore this estimate varies as the samples drawn from the simulation change. In this work, we capture this variability at each design point via the intrinsic noise in SK.
Previously, to construct the global surrogate SK model, \cite{chen2016efficient} estimated $\mathrm{CVaR}$ and its variance empirically.  Note that equation \eqref{eq:emp} can be rewritten as: 
\begin{equation}
    \widehat{CVaR_\alpha}(X)=
    \widehat{VaR_\alpha} + \dfrac{1}{n(1-\alpha)} \sum_{i = 1}^{n}(X_i - \widehat{VaR_\alpha} )^+.
    \label{eq:emp2}
\end{equation}
Therefore, an estimator of variance \citep{trindade2007financial} can be  constructed as 
\begin{equation}
    \widehat{V} = \dfrac{1}{n(n-1)}\sum^{n}_{i=1}(W_i - \Bar{W})^2,
    \label{eq:estvaremp}
\end{equation}
where $    W_i=
    \widehat{VaR_\alpha} + \dfrac{1}{1-\alpha} (X_i - \widehat{VaR_\alpha} )^+ 
$ and  $\bar{W} = \frac{1}{n}\sum_{i=1}^{n} W_i$ is equivalent to $\widehat{CVaR_\alpha}$. 

In  \cite{9384101} the authors first considered the same approach of estimating $\mathrm{CVaR}$ using the POT, however, the variance of $\mathrm{CVaR}$ (i.e. expectation of the squared deviation of $\mathrm{CVaR}$ observations from its mean) was computed by replicating the estimation of $\mathrm{CVaR}$ multiple times at each design point \eqref{eq:var-emp}.

\blue{In this work, we propose an estimator for the variance of $\mathrm{CVaR}$ that directly uses  EVT. Compared to \citet{9384101}, this reduces the computational effort by the factor of $n$, where $n$ is the number of replications at each design point (refer to Table \ref{tab:exp}). Furthermore, it provides a fair comparison between the parametric and non-parametric estimation of $\mathrm{CVaR}$ and its variance, and consequently, the performance of the SK model obtained using the POT or empirical $\mathrm{CVaR}$ estimation. The next subsection details the proposed approach.}

\subsection{POT-bsed estimation of $\mathrm{CVaR}$ variance}
Recall that  we use MLE method to estimate the $\mathrm{GPD}$ parameters $\widehat{\xi}, \widehat{\beta}$. Given the parameter values, we then obtain a Gaussian asymptotic confidence interval for $\mathrm{CVaR}$ at each design point based on equation \eqref{eq:var-cvar}. \cite{troop2019risk} first proposed an approximate asymptotic confidence interval for the $\mathrm{CVaR}$
estimate. Assuming that the approximation of the tail distribution by a $\mathrm{GPD}$ is exact, using the delta method, the variance of $CVaR_\alpha(X)$  is given by

\begin{equation}
    \widehat{V}[h(\xi, \beta)] \approx \frac{1}{N_u} \Big [  \nabla{h}(\widehat{\xi}, \widehat{\beta})\Big]^T   \mathcal{I}^{-1}\Big [ \nabla{h}(\widehat{\xi}, \widehat{\beta})\Big],
    \label{eq:var-cvar}
\end{equation}
where $N_u$ is the number of exceedances, $\mathcal{I}^{-1}$ is the inverse of the Fisher information matrix, and $\nabla h(\xi, \beta)$ 
is the column vector representing the gradients of $h$, i.e., 
$$ h(\xi, \beta) = {q} + \frac{\beta - {\xi}(q -u)}{1-{\xi}}.$$

The partial derivative of $h(\xi, \beta)$ can then be obtained as 
$$\dfrac{\partial h(\xi, \beta)}{\partial \xi} = \dfrac{q - u + \beta}{(1-\xi)^2} ,  \quad  \dfrac{\partial h(\xi, \beta)}{\partial \beta} = \dfrac{1}{1-\xi}. $$
Consequently, the Fisher information matrix \citep[see][for more details]{troop2019risk} then can be  evaluated  as follows:
\begin{equation}
\mathcal{I} \equiv - \mathbbm{E}\begin{bmatrix}
   \dfrac{\partial^2 \log \ g_{\xi, \beta}(Z)}{\partial \xi^2}      & \dfrac{\partial^2 \log \ g_{\xi, \beta}(Z)}{\partial \xi \partial \beta}   \\
    \dfrac{\partial^2 \log \ g_{\xi, \beta}(Z)}{\partial \xi \partial \beta}       & \dfrac{\partial^2 \log \ g_{\xi, \beta}(Z)}{\partial \beta^2}  \\
\end{bmatrix} \approx - \dfrac{1}{N_u}\sum_{j= 1}^{N_u}\begin{bmatrix}
   \dfrac{\partial^2 log \ g_{\xi, \beta}(z_j)}{\partial \xi^2}      & \dfrac{\partial^2 log \ g_{\xi, \beta}(z_j)}{\partial \xi \partial \beta}   \\
    \dfrac{\partial^2 \log \ g_{\xi, \beta}(z_j)}{\partial \xi \partial \beta}       & \dfrac{\partial^2 \log \ g_{\xi, \beta}(z_j)}{\partial \beta^2} \\
\end{bmatrix},
\end{equation}
where $Z$ is a random variable with a $GPD(\xi, \beta)$ distribution.

We can then find partial derivatives from the information matrix for the case $\xi \neq 0$  as,
\begin{align*}
\dfrac{\partial}{\partial \beta} \log \ g_{\xi, \beta}(z) &= \dfrac{1}{\beta} \Big[ \dfrac{z(\xi + 1)}{\beta + \xi z} - 1 \Big ]\\
\dfrac{\partial}{\partial \xi} \log \ g_{\xi, \beta}(z) &= \dfrac{1}{\xi^2} log \Big( 1 + \dfrac{\xi z}{\beta}\Big ) - \Big( \dfrac{1}{\xi} + 1\Big) \dfrac{z}{\beta + \xi z}\\
\dfrac{\partial^2}{\partial \beta^2} \log \ g_{\xi, \beta}(z) &= -\dfrac{1}{\beta^2} \Big[ \dfrac{z(\xi+ 1)}{\beta + \xi z} - 1\Big ] - \Big[ \dfrac{z(\xi+ 1)}{\beta(\beta + \xi z)^2} \Big ]\\
\dfrac{\partial^2}{\partial \beta \partial \xi} log \ g_{\xi, \beta}(z) &= \dfrac{1}{\beta} \Big[ \dfrac{z}{\beta + \xi z} - \dfrac{z^2(\xi + 1)}{(\beta + \xi z)^2} \Big ]\\
\dfrac{\partial^2}{\partial \xi^2} \log \ g_{\xi, \beta}(z) & = -\dfrac{2}{\xi^3} \log \Big( 1 +  \dfrac{\xi z}{\beta} \Big ) + \dfrac{2}{\xi^2} \dfrac{z}{\beta + \xi z} + \Big (  \dfrac{1}{\xi} + 1\Big ) \dfrac{z^2}{(\beta + \xi z)^2}.
\end{align*}
\blue{Putting it all together, equation \eqref{eq:var-cvar} then provides the desired estimate for variance of $\mathrm{CVaR}$.}

Once a sample of observations is drawn for each design point, using the definitions, $\mathrm{CVaR}$ and the estimate of the variance of $\mathrm{CVaR}$ can be calculated. 
 Further, the set of the locations of the design points, the estimated $\mathrm{CVaR}$, and its variance at each design point $(x^{[i]}, CVaR_\alpha^{[i]}, \widehat{V}^{[i]})$ are used to construct the SK model. Table \ref{tab:methods} summarizes the three different approaches to estimating the variance of  $\mathrm{CVaR}$ at each design point, where we outline the main characteristics, the corresponding equation, and whether it applies to the empirical or POT estimate of $\mathrm{CVaR}$ itself.

\begin{table}[!h]
\centering
\resizebox{.8\textwidth}{!}{%
\begin{tabular}{|l|c|p{5cm}|l|}
\hline
Method                          & Equation & Comments & Estimate of CVaR \\ \hline
Expectation of the squared deviation &  \eqref{eq:var-emp} & used in  \cite{9384101} & POT or empirical \\
\hline
Empirical                       &  \eqref{eq:estvaremp}      &  first proposed by \citep{trindade2007financial} and  used in \citep{chen2016efficient}    & empirical        \\ 
\hline
Approximate asymptotic variance &  \eqref{eq:var-cvar}    &   proposed in general in \cite{troop2019risk}. We propose to use this estimate in SK as an intrinsic variance  estimate & POT      \\
\hline
\end{tabular}%
}
\caption{Summary of the methods for estimating the variance of $\mathrm{CVaR}$ at each design point}
\label{tab:methods}
\end{table}

In the metamodeling phase, the first step is estimating the parameters of the metamodel by maximizing the likelihood function  \eqref{eq:likelihood} simultaneously over the parameters of the kernel. The metamodel is then validated over the test set. If the performance is satisfactory, the model can be used globally over the domain to quantify the risk measure at any location. If the performance is not satisfactory, the whole process can be repeated, or a larger number of design points can be used to improve the overall performance of the metamodel. 

\subsection{Generalization to other coherent measures of risk}

While the derivation above is explicitly constructed for $\mathrm{CVaR}$, it can be observed that, in principle, similar construction is possible for any coherent measure of risk subject to some non-restrictive assumptions.  To that end, recall that \cite{acerbi2002spectral} proposed   the following spectral representation for a  coherent measure of risk,
\begin{equation}
    \mathcal{R}(X) = \int_{0}^{1} VaR_\lambda (X) \, \phi(\lambda) \, \dif \lambda,
    \label{eq:cohmeas}
\end{equation}
where $\phi$ is an admissible risk spectrum, i.e.,  $\phi \in \mathcal{L}^1([0, 1])$
is positive, non-decreasing, and integrating to 1. It can further be shown that this representation is complete for the set of coherent measures of risk that are also law-invariant
and comonotonic additive, see also \citet{kusuoka2001law} for more details. 

Note that while not all coherent measures of risk are law-invariant and/or comonotone additive, the two properties are natural for most  practical use cases. 
Law-invariance requires the risk measure to give the same value of risk for all random variables that have the same distribution. Recall that two random variables $X$ and $Y$ are comonotone, if there exist increasing real functions $f$ and $g$ and another random variable $Z$, such that $X=f(Z)$ and $Y=g(Z)$. This implies that the two cannot serve as a hedge to each other, and therefore, a risk measure can  be expected to be comonotone additive, since risk should not be reduced by combining two comonotone outcomes together, as no diversification is possible. Consequently, both properties are often considered essential for any practical measure of risk. In other words, coherent measures of risk that allow for the spectral representation above span all risk measures of practical interest. 
See \citet{krokhmal2013modeling} and references therein for more details on the interpretation of these and other properties of risk measures.

 Using equation \eqref{eq:VaR}, $VaR_{\alpha}(X)$ can be replaced with its closed-form formula. Therefore, we can estimate the coherent risk measure using POT for any $\phi$ as follows. 
\begin{align}
\mathcal{R}(X) \approx h(\xi, \beta) &= \int_{0}^{1} \left (u + \frac{\beta}{\xi} \Big[ 
(\frac{\lambda}{F(u)})^{-\xi} - 1
\Big] \right ) \, \phi(\lambda) \, \dif \lambda \\
 &= u - \frac{\beta}{\xi} + \frac{\beta \,F(u)^\xi}{\xi} \int_{0}^{1} (\lambda)^{-\xi}\, \phi(\lambda) \, \dif \lambda.\label{eq:h}
\end{align}

Note that in general, the above integral can be improper at the lower boundary, however, for a proper risk measure the integral will
always be well-defined and finite \cite{acerbi2002spectral}. For instance, for  the case of $\mathrm{CVaR}$, 
\begin{equation}
    CVaR_{\alpha}(X) = \frac{1}{1-\alpha}\int_{\alpha}^{1}VaR_{\lambda}\, \dif \lambda.
\end{equation}

Following an analogous approach to \citep{troop2019risk}, we can estimate the variance of any coherent risk measure achieved from equation \eqref{eq:cohmeas}. To do so, the $\nabla h(\xi, \beta)$ 
which is the column vector representing the gradients of $h(\xi, \beta)$, should be derived.
While an analytical expression may not be possible to obtain in general, for any given measure, i.e., any given function $\phi$, such a derivation can follow the same approach as presented in the previous section in the  case of $\mathrm{CVaR}$. Even if such a derivation is not possible, equation \eqref{eq:h} can be differentiated numerically to produce the required estimates. 
%

It is also worth noting that, while this discussion provides a pathway to estimate variance for (law-invariant and comonotone additive) coherent measures of risk in general, for any particular measure of interest a more computationally efficient approach may be possible, depending on how the measure is defined. For example, consider the Higher-Moment Coherent Measure of Risk (HMCR), proposed in \citet{krokhmal2007higher}, which generalizes $\mathrm{CVaR}$. One way to define it is by solving equation
\begin{equation}
    (1-\alpha)^{-1/(1-p)} = \frac{\norm{\big (X - \eta_{p, \alpha}(X)\big )^+}_p \hfill}{\norm{\big (X - \eta_{p, \alpha}(X)\big )^+}_{p-1}} , \quad p > 1.
\end{equation}
Its solution $\eta_{p, \alpha}(X)$ plays the role of VaR in $\mathrm{CVaR}$ definition, so, for example, Second-Moment Coherent Measure of Risk (SMCR) is given as 

\begin{align}
    SMCR_{\alpha}(X) &= \eta_{2, \alpha}(X) + (1-\alpha)^{-2}\norm{(X - \eta_{2, \alpha}(X))^+}_1 \\
    &= \eta_{2, \alpha}(X) + (1-\alpha)^{-2}E{(X - \eta_{2, \alpha}(X))^+}
\end{align}
This expression then can be used directly in the same way as the VaR-based definition of $\mathrm{CVaR}$ to construct POT estimate. 


\section{Experiments and Results}
\label{experandres}

\subsection{Experimental Design}


We develop two test cases to demonstrate our methodology's capabilities. The first case, similar to \citep{9384101}, involves a benchmark function studied in three scenarios, introducing different random noise types (triangular, Pareto, normal distributions) to create a complex surface affecting $\mathrm{CVaR}$ estimation. The second case, inspired by \cite{chen2016efficient}, uses a stochastic activity network to illustrate practical application in a complex simulation model.

The first test case is a  widely-used two-dimensional function, given as $	    f_i(x_1,x_2) = x_1\sin{(\pi x_2)}+x_2\sin{(\pi x_1)} + \epsilon_i (x_1,x_2)$,
where $\epsilon_i (x_1,x_2)$ is random noise and $x_1, x_2 \in [-\pi,\pi]$. The random noise is studied under three different scenarios as described in Table \ref{tab:cvar-noise}. The function, as well as true $\mathrm{CVaR}$ values, are as discussed in detail  in \citet{norton2019calculating}. The True $\mathrm{CVaR}$ column in Table \ref{tab:cvar-noise}, shows the analytical closed-form formula for computing $\mathrm{CVaR}$ under the three different scenarios.

\begin{table}[!h]
\centering
\normalsize 
\caption{Additive noise terms and expressions for calculating the true $\mathrm{CVaR}$ values analytically}
\label{tab:cvar-noise}

\begin{tabular}{|l|l|l|}
\hline
Scenario Description & $\epsilon_i(x_1,x_2)$                     & True $\mathrm{CVaR}$ $[\epsilon_i(x_1,x_2)]$                                              \\ \hline
normal (thin) tail   & $N(\mu=0,\sigma=\sqrt{(x_1^2+x_2^2)})$      & $\frac{\sqrt{(x_1^2+x_2^2)}\phi(\Phi^{-1} (\alpha))}{\alpha}$                 \\ \hline
finite tail          & $Triangular(\min=0,\max= \sqrt{(x_1^2+x_2^2)})$   & $\sqrt{(x_1^2+x_2^2)} \Big ( 1-  \frac{\sqrt{2(1-\alpha)}}{3} \Big )$ \\ \hline
heavy tail           & $Pareto(a=2, x_m=2+\sqrt{(x_1^2 +x_2^2)}) $ & $\frac {2(2+\sqrt{(x_1^2+x_2^2)} )}{(1-\alpha)^{1/2}}$                        \\ \hline
\end{tabular}
\end{table}

Multiple observations of the simulation response at each design point are required (denoted by $N$ in this work) to estimate $\mathrm{CVaR}$. Moreover, to empirically evaluate the intrinsic variance using equation \eqref{eq:var-emp}, it is required to replicate the estimation of $\mathrm{CVaR}$. The number of replications at each design point is denoted by $n$. Note that the proposed approach does not require multiple replications. Finally, the SK model is constructed by sampling observations from $K$ design points. 
Experimental trials are designed to assess the effects of the different components of this framework over the final performance of the metamodel for different combinations of $N, k, n$. Table \ref{tab:exp} presents the different scenarios and computational budget levels. Naturally, the total computational budget (measured in the number of evaluations) can be calculated as $N\times k \times n$.
We consider three values for the total computational budget (number of function evaluations): $10^{5}, 10^6$, and $ 10^{7}$. The higher budget should directly result in a more accurate estimation. Note though, that since this improvement happens across all models and our main goal is to compare relative performance, we do not consider more budget levels. Further, observe that the tail parameter $\alpha$ is closely connected to the total budget (the higher the value of $\alpha$ the fewer observations are included in the tail), and hence varying $\alpha$ is closely tied to varying the budget.

For each value of the total budget, we consider a variety of possible allocations between the number of design points, observations, and replications, as given in Table \ref{tab:exp}. The higher the value of each of the parameters, the more accurate the resulting model. On the other hand, since the three are constrained by the total budget, it may be interesting to evaluate which combination renders the best performance.  Further, we expect that there may be a difference in the relative performance of the different models for different combinations. Consequently, in the remainder of this section, when discussing the results, we both compare all cases for each of the three budgets and analyze the relative performance between different models within each case.


\begin{table}[!h]
\centering
\caption{Budget breakdown and test cases for the set of experiments with benchmark function}
\label{tab:exp}
\resizebox{\linewidth}{!}{%
\begin{tabular}{|c|r|r|r|c|}
\hline
Budget Identifier & Number of design points ($k$) & Number of replications ($n$) & Number of Observations ($N$) & Total Budget  \\
\hline

1 & 50 \hspace{2cm}   & 10 \hspace{2cm}  & 200 \hspace{2cm}  & \multirow{5}{*}{\rotatebox[origin=c]{90}{100,000}}\\

2 & 50 \hspace{2cm}   & 5 \hspace{2cm}  & 400 \hspace{2cm}  & \multirow{5}{*}{\rotatebox[origin=c]{90}{}}\\

3 & 50 \hspace{2cm}   & 1 \hspace{2cm}  & 2000 \hspace{2cm}  & \multirow{5}{*}{\rotatebox[origin=c]{90}{}}\\

4 & 100 \hspace{2cm}   & 5 \hspace{2cm}  & 200 \hspace{2cm}  & \multirow{5}{*}{\rotatebox[origin=c]{90}{}}\\

5 & 100 \hspace{2cm}   & 1 \hspace{2cm}  & 1000 \hspace{2cm}  & \multirow{5}{*}{\rotatebox[origin=c]{90}{}}\\

\hline
\hline

6 & 50 \hspace{2cm}   & 20 \hspace{2cm}  & 1000 \hspace{2cm}  & \multirow{5}{*}{\rotatebox[origin=c]{90}{1,000,000}}\\

7 & 50 \hspace{2cm}   & 10 \hspace{2cm}  & 2000 \hspace{2cm}  & \multirow{4}{*}{}\\

8 & 100 \hspace{2cm} & 10 \hspace{2cm} & 1000 \hspace{2cm} &  \\

9 & 100 \hspace{2cm} & 5  \hspace{2cm} & 2000  \hspace{2cm} &  \\

10 & 100 \hspace{2cm} & 1 \hspace{2cm}  & 10,000 \hspace{2cm} & \\ \hline 
\hline

11 & 50 \hspace{2cm} & 100\hspace{2cm} & 2000 \hspace{2cm}  & \multirow{5}{*}{\rotatebox[origin=c]{90}{10,000,000}} \\

12 & 100 \hspace{2cm} & 50 \hspace{2cm}  & 2000 \hspace{2cm}  &  \\

13 & 100 \hspace{2cm} & 10 \hspace{2cm}  & 10,000 \hspace{2cm}  &  \\

14 & 100 \hspace{2cm} & 5 \hspace{2cm}   & 20,000 \hspace{2cm} &  \\

15 & 100 \hspace{2cm} & 1 \hspace{2cm}   & 100,000 \hspace{2cm} & \\
\hline
\end{tabular}%

}

\end{table}

The cases with $n = 1$ are designed to evaluate the performance of the meta-model where the intrinsic variance is estimated using one replication only. This shows the potential improvement of assigning all the budget to the observations (i.e. estimating $\mathrm{CVaR}$ once), instead of replicating the $\mathrm{CVaR}$ estimate $n$ times and consequently estimating the empirical variance using equation \eqref{eq:var-emp}.

 In the second set of experiments we evaluate our method with a simple Stochastic Activity Network (SAN) considered by \cite{chen2016efficient}. Five
activities are involved in the completion of a project. The time to finish activity $i$ is denoted by $T_i$ and follows an exponential distribution. The activity
times $T_i$ are assumed to be i.i.d of the other activity times. $T_3$
follows an exponential distribution with mean $x \in [0.3, 2]$, while for any $i \neq 3$, $T_i$ is exponentially distributed with mean 1. In this network, three paths are defined. The project completion time is specified as follows: 
$$L(x) = \max\{T_1 + T_2, T_1 + T_3(x), T_4 + T_5\}$$
We are then interested in estimating $\mathrm{CVaR}$ as a function of $x$. The experimental design space for the mean parameter $x$ is $x \in [0.3, 2]$. Seven equally spaced design points are chosen from the design space. At each design point, $N$ observations are simulated. Based on these observations,  $\mathrm{CVaR}$ and its variance are estimated.

Choosing an appropriate spatial correlation function is one of the most important steps in Gaussian processes. In this study a Gaussian correlation function is used, given by 
\[r(x-x^\prime,\theta )= exp\Big(-\sum_i\theta_i(x_i-x_i^\prime)^2\Big)\] where $\theta_i$ determines how quickly the spatial correlation decays as
the two points become farther apart in each dimension $i$. Finding a well-specified polynomial trend can be burdensome and might result in the ill-specification of the model as the test function is designed to have many oscillations. Therefore, Stochastic kriging is performed with $f(x)^T \beta = \beta_0$ which is widely used in practice due to its simplicity.  However, it is worth noting that SK without the trend terms can still lead to good predictions if the kernel and its parameters are well-estimated.

The design points are selected with a Latin hypercube sampling (LHS) procedure, one form of stratified sampling to generate a near-random sample of parameter values in a multidimensional domain. In the one-dimensional case, the CDF is divided into $n$ equal partitions, and a random point is sampled from each partition. In the multidimensional case, a one-dimensional LHS sample is drawn for each dimension, which is then combined randomly. This design is often used to construct computer experiments to reduce computational effort. Theoretical results demonstrate that LHS shows favorable performance, even in multidimensional cases, see for example, \citet{helton2003latin}


Mean Average Percentage Error (MAPE) is used to evaluate the performance of the surrogate models. 
In order to draw fair comparisons between the different noise distributions, as well as between  different simulation budgets/cases, we evaluate and test the metamodels constructed for each budget over the same test set which is sampled using LHS across the domain, independently from the set of design points.

To assess the performance of our proposed method denoted by POT-EVT (i.e., POT estimate of $\mathrm{CVaR}$ and EVT estimate of its variance), the results of three other approaches are used as benchmarks.
\begin{itemize}
    \item Ordinary kriging where we ignore the intrinsic noise in the estimations of $\mathrm{CVaR}$. As was expected, in the cases where the number of observations is relatively large, the estimate of $\mathrm{CVaR}$ becomes very accurate (i.e very close to the true $\mathrm{CVaR}$) and therefore, the ordinary kriging performs as well as the other methods. This method is labeled as ORD-KRG. 
    
    \item A global SK with POT estimate of $\mathrm{CVaR}$ and intrinsic variance as the expectation of the squared deviation of $\mathrm{CVaR}$ estimates  \eqref{eq:var-emp} at each design point. This method builds on the previous approach outlined in \cite{9384101} and is denoted by POT-EMP in our results (i.e., POT $\mathrm{CVaR}$ estimation with the empirical estimation of variance). 
    
    \item An SK model with empirical estimates of both $\mathrm{CVaR}$ and its variance. Note that this method was considered by \cite{chen2016efficient}. The variance of $\mathrm{CVaR}$ is estimated based on \eqref{eq:estvaremp}. This method is denoted by EMP-EMP in our results. 
    
\end{itemize}

\blue{Note that  EMP-EMP and our proposed approach, POT-EVT, only require $n=1$. For the cases with the number of replications, $n$, greater than 1, each associated replication with a given design point would have its own variance estimate. The mean and variance of the sample mean, $\bar{Y}$, are used to construct the kriging models. That is, at each design point, the mean of $\mathrm{CVaR}$ estimations, $\hat{E}(\bar{Y})$, and the variance of the mean, $\hat{Var}(\bar{Y})$ 
(equations \eqref{eq:cvar_mean}, \eqref{eq:cvar_var} respectively) are used as the replications are independent of one another, i.e., 
\begin{equation}
    \hat{E}(\bar{Y}) = \frac{CVaR_{1} + CVaR_{2} + 
 \dots+CVaR_{n}}{n}
 \label{eq:cvar_mean}
\end{equation}
\begin{equation}
\hat{Var}(\bar{Y}) =  \frac{Var(CVaR_{1})}{n^2} + \frac{Var(CVaR_{2})}{n^2} + \dots + \frac{Var(CVaR_{n})}{n^2} 
 \label{eq:cvar_var}
\end{equation}
}

Each combination of the budget allocation $(k-n-N)$ is replicated 10 times and the median of the performance of the constructed kriging models is reported in the results section.

\subsection{Experimental Results}


\subsubsection{Benchmark function}
\begin{figure}[!htb]
     \centering
     \includegraphics[width=.99\linewidth]{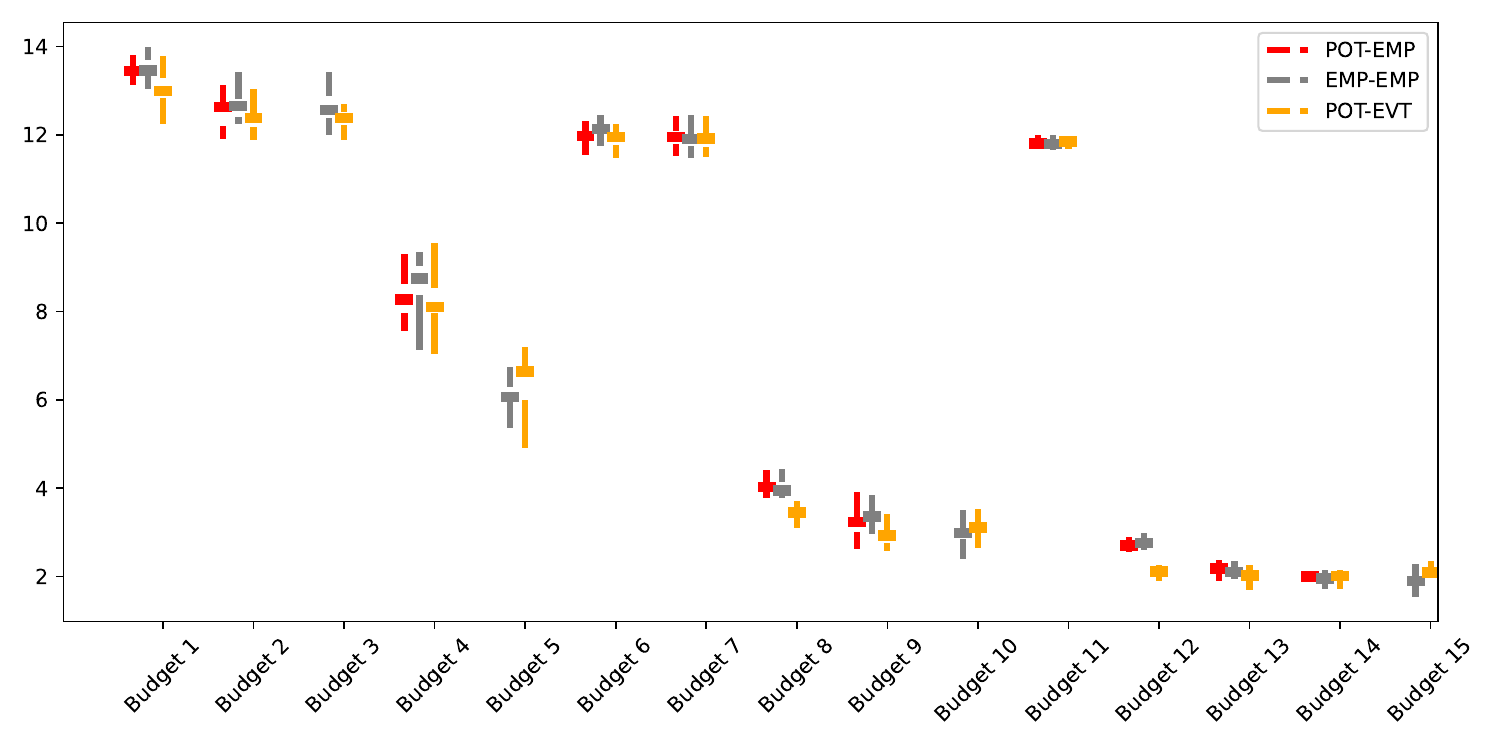}
     \caption{Boxplots for the normal noise $\alpha = 0.99 $ }\label{Fig:normal99}
\end{figure}

\begin{figure}[!htb]

     \centering
     \includegraphics[width=.99\linewidth]{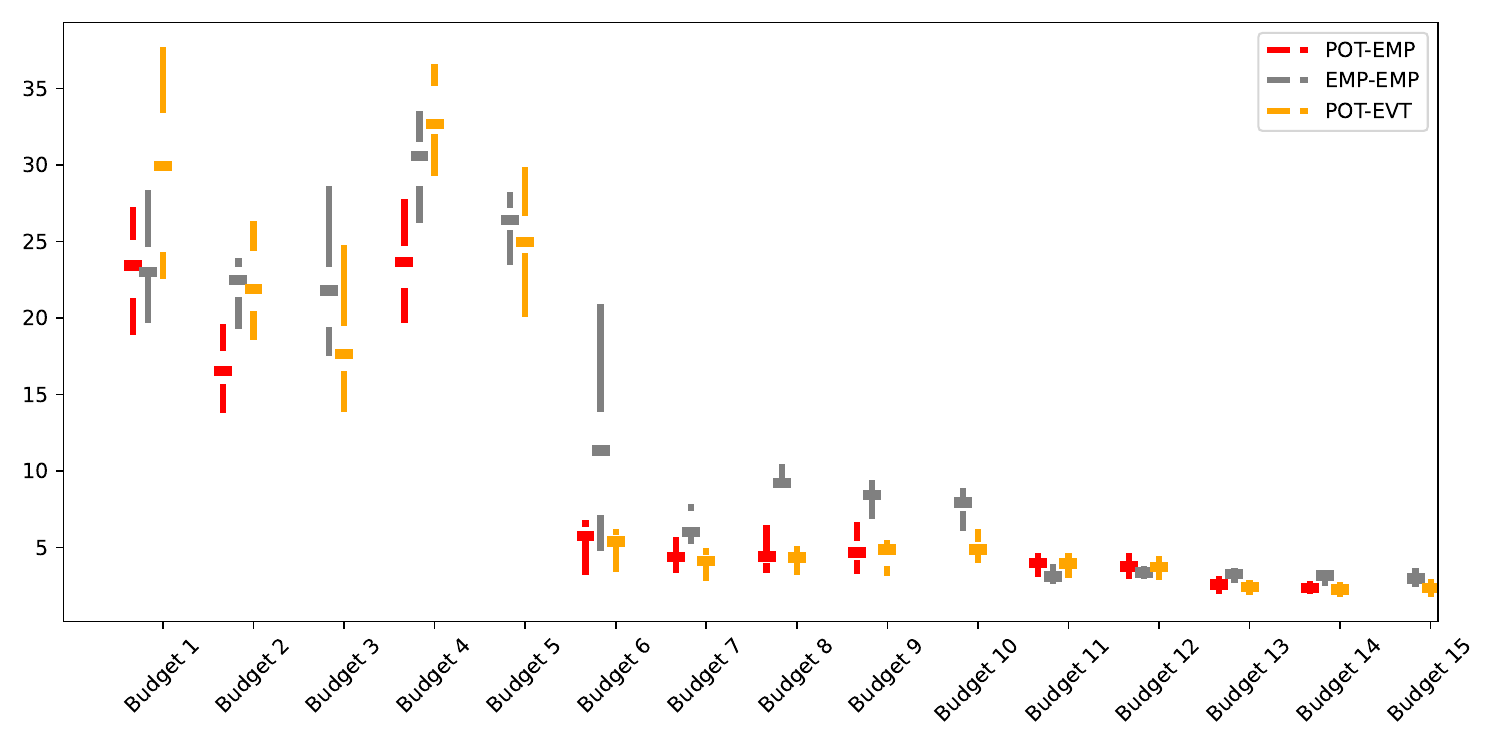}
     \caption{Boxplots for the Pareto noise $\alpha = 0.99 $}\label{Fig:pareto99}
\end{figure}

\begin{figure}[!h]
    \centering
    \includegraphics[width =.99\linewidth]{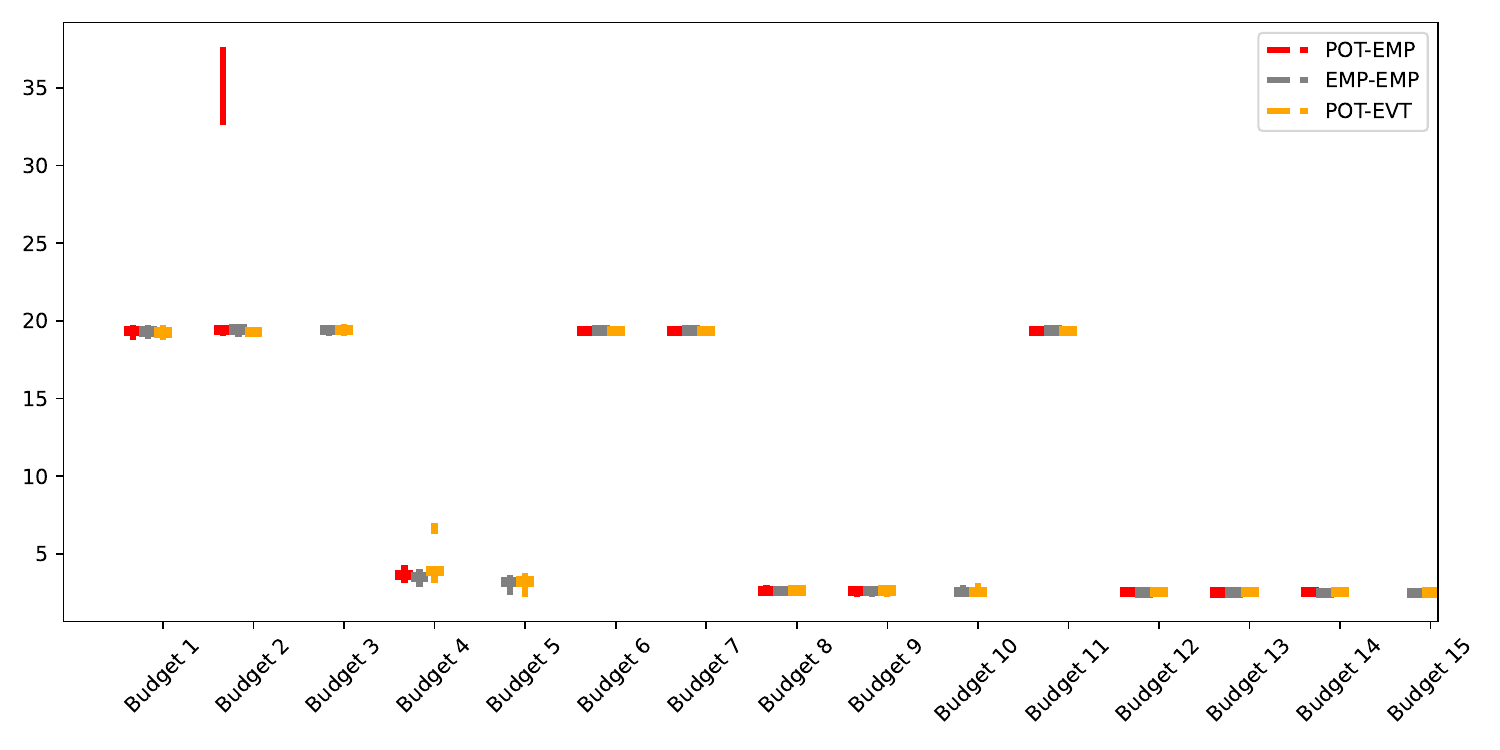}
    \caption{Boxplots for the triangular noise $\alpha = 0.99 $}
    \label{fig:triang99}
\end{figure}

Tables \ref{tab:firstbudget}--\ref{tab:pval}, and Figures \ref{Fig:normal99}--\ref{fig:triang99} summarise the results for the first set of experiments with the benchmark function. Specifically, Tables \ref{tab:firstbudget}--\ref{tab:thirdbudget} report the median MAPE for the four methods across the three budget levels. The same data is depicted in Figures \ref{Fig:normal99}-- \ref{fig:triang99} as boxplots (based on the 10 replications of the estimates). Note that here we only present the case of $\alpha=0.99$, while the rest are given in the Appendix. Finally, Table \ref{tab:pval}  shows the results of the Wilcoxon Signed Rank test for the difference between our proposed and the EMP-EMP method.


Overall, for triangular noise, methods show similar performance, especially with larger budgets. This is expected as the triangular distribution is bounded, and with enough observations, all methods converge to estimates with low error. Additionally, our methodology is best suited for Pareto noise due to its heavy-tailed nature, which concentrates risk in the distribution tail, making accurate estimation crucial. Notably, ordinary kriging is not competitive with lower budgets (higher $\alpha$), so we do not discuss it in detail below.

As can be expected, the surrogate model fitted with a higher number of design points performs better across the different noise distributions. Table \ref{tab:firstbudget} shows the results of the test function with the three different additive noises for three different $\alpha$ levels. As expected, in the cases with 100 design points, the MAPE measures are relatively small especially for the normal and triangular noise, since the $\mathrm{CVaR}$ estimates at each design point are very close to the true $\mathrm{CVaR}$ due to the limited tail of distributions. In almost all cases, the models that take into account intrinsic variance outperform the ordinary kriging. Note that the trials with one replication (i.e. $n = 1$) assign all of the computational budgets to $\mathrm{CVaR}$ estimation, therefore the $\mathrm{CVaR}$ estimates are very accurate.


At each budget level ($10^5, 10^6$, and $10^7$), comparing different trials reveals the trade-off between design points, replications, and observations. For example, comparing 100-1-1000 and 100-5-200 demonstrates that one replication with more observations yields slightly better models than five replications with fewer observations. This suggests that using our proposed method, POT-EVT, or the EMP-EMP method, which estimates $\mathrm{CVaR}$ variance based on a single replication, outperforms allocating part of the simulation budget to replicate $\mathrm{CVaR}$ estimation for intrinsic variance calculation.


Due to the heavy-tailed nature of the Pareto distribution, $\mathrm{CVaR}$ estimates are less accurate compared to other additive noise distributions. Consequently, this leads to a larger deviation from the true $\mathrm{CVaR}$ and a higher MAPE when compared to normal noise. As anticipated, when using an $\alpha$ level of 0.95, $\mathrm{CVaR}$ estimations become relatively more accurate, thanks to a larger number of observed exceedances over the value at risk. This results in smaller variance in $\mathrm{CVaR}$ estimations, improving the overall prediction of $\mathrm{CVaR}$ at this $\alpha$ level and budget.

\begin{table}[]
\centering
\caption{Median of 10 MAPEs for predicting $\mathrm{CVaR}$ with a total budget of $10^5$. Note that the second row of the table corresponds to the confidence level of the $CVaR_{\alpha}$ and the budget column shows the budget breakdown $(k-n-N)$ and the number beneath it corresponds to the budget identifier from table \ref{tab:exp}  }
\label{tab:firstbudget}
\resizebox{\textwidth}{!}{%
\begin{tabular}{|c|c|ccc|ccc|ccc|}
\hline
\multirow{2}{*}{Method} & \multirow{2}{*}{Budget*}                                                       & \multicolumn{3}{c|}{Normal}                                                      & \multicolumn{3}{c|}{Pareto}                                                      & \multicolumn{3}{c|}{Triangular}                                                  \\ \cline{3-11} 
                        &                                                                         & \multicolumn{1}{c}{0.95} & \multicolumn{1}{c}{0.99} & \multicolumn{1}{c|}{0.995} & \multicolumn{1}{c}{0.95} & \multicolumn{1}{c}{0.99} & \multicolumn{1}{c|}{0.995} & \multicolumn{1}{c}{0.95} & \multicolumn{1}{c}{0.99} & \multicolumn{1}{c|}{0.995} \\ \hline
ORD-KRG                 & \multirow{4}{*}{\rotatebox[origin=c]{90}{\begin{tabular}[c]{@{}c@{}}50-10-200\\ 1\end{tabular}}} & 21.06                    & 14.48                    & 13.89                      & 42.39                    & 125.73                   & 146.70                     & 28.68                    & 24.52                    & 23.83                      \\
POT-EMP                 &                                                                         & 17.28                    & 13.44                    & 15.08                      & 8.18                     & 23.43                    & 27.27                      & 41.43                    & 19.37                    & 18.82                      \\
EMP-EMP                 &                                                                         & 17.47                    & 13.46                    & 14.78                      & 8.10                     & 23.01                    & 36.54                      & 31.66                    & 19.31                    & 18.77                      \\
POT-EVT                 &                                                                         & 17.15                    & 13                       & 14.35                      & 15.56                    & 29.92                    & 34.08                      & 22.79                    & 19.25                    & 18.82                      \\ \hline
ORD-KRG                 & \multirow{4}{*}{\rotatebox[origin=c]{90}{\begin{tabular}[c]{@{}c@{}}50-5-400\\ 2\end{tabular}}}                                                       & 20.99                    & 14.14                    & 13.10                      & 35.02                    & 69.37                    & 71.68                      & 28.75                    & 24.65                    & 23.86                      \\
POT-EMP                 &                                                                         & 17.02                    & 12.63                    & 11.75                      & 7.75                     & 16.53                    & 21.80                      & 41.52                    & 19.4                     & 18.84                      \\
EMP-EMP                 &                                                                         & 17.08                    & 12.66                    & 11.94                      & 7.64                     & 22.46                    & 30.42                      & 41.47                    & 19.44                    & 18.82                      \\
POT-EVT                 &                                                                         & 17.14                    & 12.38                    & 11.49                      & 10.90                    & 21.88                    & 28.43                      & 21.89                    & 19.29                    & 18.81                      \\ \hline
ORD-KRG                 & \multirow{4}{*}{\rotatebox[origin=c]{90}{\begin{tabular}[c]{@{}c@{}}50-1-2000\\ 3\end{tabular}}}                                                       & 21.11                    & 14.10                    & 12.88                      & 10.91                    & 16.91                    & 21.31                      & 28.76                    & 24.69                    & 23.95                      \\
POT-EMP                 &                                                                         & -                        & -                        & -                          & -           & -            & -                &     -                     &       -                   &    -                        \\
EMP-EMP                 &                                                                         & 17.14                    & 12.57                    & 11.64                      & 8.58                     & 21.80                    & 28.72                      & 41.65                    & 19.43                    & 18.93                      \\
POT-EVT                 &                                                                         & 17.05                    & 12.39                    & 12.05                      & 8.29                     & 17.64                    & 23.66                      & 21.87                    & 19.41                    & 18.89                      \\ \hline
ORD-KRG                 & \multirow{4}{*}{\rotatebox[origin=c]{90}{\begin{tabular}[c]{@{}c@{}}100-5-200\\ 4\end{tabular}}}                                                       & 7.38                     & 9.27                     & 10.65                      & 49.42                    & 99.78                    & 158.97                     & 3.79                     & 3.91                     & 4.19                       \\
POT-EMP                 &                                                                         & 6.29                     & 8.27                     & 10.32                      & 10.46                    & 23.66                    & 33.22                      & 3.40                     & 3.66                     & 4                          \\
EMP-EMP                 &                                                                         & 6.7                      & 8.75                     & 10.53                      & 11.40                    & 30.56                    & 41.84                      & 3.33                     & 3.50                     & 3.53                       \\
POT-EVT                 &                                                                         & 5.77                     & 8.1                      & 9.83                       & 14.29                    & 32.69                    & 41.76                      & 6.88                     & 3.92                     & 4.22                       \\ \hline
ORD-KRG                 & \multirow{4}{*}{\rotatebox[origin=c]{90}{\begin{tabular}[c]{@{}c@{}}100-1-1000\\ 5\end{tabular}}}                                                       & 6.79                     & 7.59                     & 8.5                        & 15.98                    & 26.55                    & 30.81                      & 3.64                     & 3.39                     & 3.44                       \\
POT-EMP                 &                                                                         & -                        & -                        & -                          & -                        & -                        & -                          & -                         &     -                     &      -                      \\
EMP-EMP                 &                                                                         & 5.88                     & 6.07                     & 7.26                       & 10.24                    & 26.40                    & 36.13                      & 3.34                     & 3.20                     & 3.13                       \\
POT-EVT                 &                                                                         & 6.29                     & 6.64                     & 7.55                       & 10.78                    & 24.96                    & 31.93                      & 3.50                     & 3.21                     & 3.20                       \\ \hline
\end{tabular}%
}

\end{table}

\begin{table}[]
\centering
\caption{Median of 10 MAPEs for predicting $\mathrm{CVaR}$ with a total budget of $10^6$}
\label{tab:secondbudget}
\resizebox{\textwidth}{!}{%
\begin{tabular}{|c|c|ccc|ccc|ccc|}
\hline
\multirow{2}{*}{Method} & \multirow{2}{*}{Budget} & \multicolumn{3}{c|}{Normal} & \multicolumn{3}{c|}{Pareto} & \multicolumn{3}{c|}{Triangular} \\ \cline{3-11} 
 &  & 0.95 & 0.99 & 0.995 & 0.95 & 0.99 & 0.995 & 0.95 & 0.99 & 0.995 \\ \hline
ORD-KRG & \multirow{4}{*}{\rotatebox[origin=c]{90}{\begin{tabular}[c]{@{}c@{}}50-20-1000\\ 6\end{tabular}}} & 21.04 & 13.87 & 12.33 & 7.61 & 14.52 & 19.43 & 28.73 & 24.71 & 23.94 \\
POT-EMP &  & 16.94 & 11.97 & 10.81 & 5.24 & 5.73 & 15.65 & 41.50 & 19.36 & 18.86 \\
EMP-EMP &  & 16.96 & 12.13 & 10.86 & 4.73 & 11.33 & 13.41 & 41.45 & 19.38 & 18.88 \\
POT-EVT &  & 16.98 & 11.96 & 10.77 & 5.04 & 5.38 & 14.27 & 21.84 & 19.36 & 18.86 \\ \hline
ORD-KRG & \multirow{4}{*}{\rotatebox[origin=c]{90}{\begin{tabular}[c]{@{}c@{}}50-10-2000\\ 7\end{tabular}}} & 21.07 & 13.78 & 12.19 & 4.40 & 7.72 & 10.15 & 28.76 & 24.71 & 23.94 \\
POT-EMP &  & 17.04 & 11.96 & 10.88 & 4.59 & 4.39 & 5.09 & 41.57 & 19.37 & 18.90 \\
EMP-EMP &  & 16.97 & 11.91 & 10.88 & 4.64 & 6.03 & 12.42 & 41.53 & 19.40 & 18.92 \\
POT-EVT &  & 17.03 & 11.92 & 10.74 & 4.54 & 4.12 & 4.52 & 21.82 & 19.37 & 18.90 \\ \hline
ORD-KRG & \multirow{4}{*}{\rotatebox[origin=c]{90}{\begin{tabular}[c]{@{}c@{}}100-10-1000\\ 8\end{tabular}}} & 3.68 & 4.10 & 4.31 & 9.77 & 16.12 & 24.78 & 3.05 & 2.84 & 2.87 \\
POT-EMP &  & 3.46 & 4.03 & 4.30 & 4.68 & 4.41 & 10.24 & 2.78 & 2.64 & 2.67 \\
EMP-EMP &  & 3.54 & 3.95 & 4.50 & 4.65 & 9.20 & 14.02 & 2.82 & 2.61 & 2.59 \\
POT-EVT &  & 3.09 & 3.45 & 3.49 & 4.56 & 4.35 & 14.69 & 2.79 & 2.64 & 2.68 \\ \hline
ORD-KRG & \multirow{4}{*}{\rotatebox[origin=c]{90}{\begin{tabular}[c]{@{}c@{}}100-5-2000\\ 9\end{tabular}}} & 3.23 & 3.47 & 3.91 & 6.13 & 10.65 & 13.14 & 3.04 & 2.81 & 2.84 \\
POT-EMP &  & 3.18 & 3.24 & 3.63 & 4.23 & 4.66 & 5.17 & 2.89 & 2.64 & 2.66 \\
EMP-EMP &  & 3.22 & 3.36 & 3.69 & 4.80 & 8.43 & 17.10 & 2.84 & 2.63 & 2.66 \\
POT-EVT &  & 3.04 & 2.93 & 3.29 & 4.48 & 4.86 & 5.00 & 2.90 & 2.66 & 2.68 \\ \hline
ORD-KRG & \multirow{4}{*}{\rotatebox[origin=c]{90}{\begin{tabular}[c]{@{}c@{}}100-1-10,000\\ 10\end{tabular}}} & 3.41 & 3.38 & 3.93 & 5.30 & 8.63 & 9.97 & 3.17 & 2.76 & 2.80 \\
POT-EMP &  & - & - & - & - & - & - & - & - & - \\
EMP-EMP &  & 2.84 & 2.99 & 3.32 & 4.81 & 7.94 & 12.63 & 2.79 & 2.54 & 2.64 \\
POT-EVT &  & 3.00 & 3.11 & 3.41 & 4.30 & 4.86 & 6.63 & 2.88 & 2.57 & 2.60 \\ \hline
\end{tabular}%
}

\end{table}

\begin{table}[]
\centering
\caption{Median for 10 MAPEs for predicting $\mathrm{CVaR}$ with a total budget of $10^7$}
\label{tab:thirdbudget}
\resizebox{\textwidth}{!}{%
\begin{tabular}{|c|c|ccc|ccc|ccc|}
\hline
\multirow{2}{*}{Method} & \multirow{2}{*}{Budget} & \multicolumn{3}{c|}{Normal} & \multicolumn{3}{c|}{Pareto} & \multicolumn{3}{c|}{Triangular} \\ \cline{3-11} 
 &  & 0.95 & 0.99 & 0.995 & 0.95 & 0.99 & 0.995 & 0.95 & 0.99 & 0.995 \\ \hline
ORD-KRG & \multirow{4}{*}{\rotatebox[origin=c]{90}{\begin{tabular}[c]{@{}c@{}}50-100-2000\\ 11\end{tabular}}} & 21.05 & 13.72 & 12.21 & 2.84 & 4.58 & 5.77 & 28.76 & 24.71 & 23.94 \\
POT-EMP &  & 16.94 & 11.81 & 10.78 & 3.33 & 3.98 & 4.73 & 41.56 & 19.36 & 18.87 \\
EMP-EMP &  & 16.93 & 11.80 & 10.84 & 3.34 & 3.11 & 3.91 & 31.67 & 19.38 & 18.89 \\
POT-EVT &  & 16.93 & 11.85 & 10.72 & 3.34 & 3.96 & 4.72 & 21.82 & 19.36 & 18.87 \\ \hline
ORD-KRG & \multirow{4}{*}{\rotatebox[origin=c]{90}{\begin{tabular}[c]{@{}c@{}}100-50-2000\\ 12\end{tabular}}} & 2.54 & 2.19 & 2.37 & 2.82 & 5.38 & 7.02 & 3.00 & 2.54 & 2.76 \\
POT-EMP &  & 2.70 & 2.70 & 3.10 & 2.27 & 3.76 & 4.33 & 2.75 & 2.56 & 2.54 \\
EMP-EMP &  & 2.73 & 2.76 & 3.19 & 2.01 & 3.35 & 4.43 & 2.75 & 2.52 & 2.51 \\
POT-EVT &  & 2.36 & 2.11 & 2.34 & 2.29 & 3.71 & 4.23 & 2.75 & 2.56 & 2.54 \\ \hline
ORD-KRG & \multirow{4}{*}{\rotatebox[origin=c]{90}{\begin{tabular}[c]{@{}c@{}}100-10-10,000\\ 13\end{tabular}}} & 2.64 & 2.22 & 2.23 & 2.04 & 3.44 & 4.12 & 2.99 & 2.68 & 2.74 \\
POT-EMP &  & 2.36 & 2.18 & 2.15 & 1.71 & 2.59 & 2.47 & 2.80 & 2.53 & 2.54 \\
EMP-EMP &  & 2.39 & 2.10 & 2.23 & 2.05 & 3.28 & 3.84 & 2.79 & 2.52 & 2.51 \\
POT-EVT &  & 2.38 & 2.02 & 2.03 & 1.69 & 2.40 & 2.43 & 2.80 & 2.54 & 2.54 \\ \hline
ORD-KRG & \multirow{4}{*}{\rotatebox[origin=c]{90}{\begin{tabular}[c]{@{}c@{}}100-5-20,000\\ 14\end{tabular}}} & 2.70 & 2.37 & 2.17 & 2.17 & 3.96 & 4.37 & 3.01 & 2.59 & 2.70 \\
POT-EMP &  & 2.42 & 2.00 & 2.02 & 1.65 & 2.36 & 2.68 & 2.78 & 2.54 & 2.48 \\
EMP-EMP &  & 2.34 & 1.96 & 2.02 & 1.85 & 3.18 & 3.93 & 2.76 & 2.49 & 2.48 \\
POT-EVT &  & 2.47 & 2.01 & 1.93 & 1.59 & 2.26 & 2.39 & 2.78 & 2.54 & 2.48 \\ \hline
ORD-KRG & \multirow{4}{*}{\rotatebox[origin=c]{90}{\begin{tabular}[c]{@{}c@{}}100-1-100,000\\ 15\end{tabular}}} & 2.63 & 2.27 & 2.26 & 2.05 & 3.27 & 4.17 & 2.94 & 2.64 & 2.74 \\
POT-EMP &  & - & - & - & - & - & - & - & - & - \\
EMP-EMP &  & 2.41 & 1.90 & 1.89 & 1.99 & 2.98 & 3.83 & 2.79 & 2.49 & 2.49 \\
POT-EVT &  & 2.43 & 2.09 & 2.14 & 1.51 & 2.36 & 2.39 & 2.79 & 2.52 & 2.47 \\ \hline
\end{tabular}%
}

\end{table}

\begin{table}[]
\centering
\caption{P-value results for the one-sided Wilcoxon Signed Rank Test over the MAPE of the POT-EVT and EMP-EMP SK models. Bolded values support our claim that our model has a lower error, while the underlined values show that the EMP-EMP method achieves a lower error.}
\label{tab:pval}
\resizebox{1.05\textwidth}{!}{%
\begin{tabular}{|c|llllll|llllll|llllll|}
\hline
\multirow{3}{*}{Budget} & \multicolumn{6}{c|}{Normal}                                                                                                                          & \multicolumn{6}{c|}{Pareto}                                                                                                                          & \multicolumn{6}{c|}{Triangular}                                                                                                                      \\ \cline{2-19} 
                        & \multicolumn{2}{c}{0.95}                        & \multicolumn{2}{c}{0.99}                        & \multicolumn{2}{c|}{0.995}                       & \multicolumn{2}{c}{0.95}                        & \multicolumn{2}{c}{0.99}                        & \multicolumn{2}{c|}{0.995}                       & \multicolumn{2}{c}{0.95}                        & \multicolumn{2}{c}{0.99}                        & \multicolumn{2}{c|}{0.995}                       \\ \cline{2-19} 
                        & \multicolumn{1}{c}{<=} & \multicolumn{1}{c}{>=} & \multicolumn{1}{c}{<=} & \multicolumn{1}{c}{>=} & \multicolumn{1}{c}{<=} & \multicolumn{1}{c|}{>=} & \multicolumn{1}{c}{<=} & \multicolumn{1}{c}{>=} & \multicolumn{1}{c}{<=} & \multicolumn{1}{c}{>=} & \multicolumn{1}{c}{<=} & \multicolumn{1}{c|}{>=} & \multicolumn{1}{c}{<=} & \multicolumn{1}{c}{>=} & \multicolumn{1}{c}{<=} & \multicolumn{1}{c}{>=} & \multicolumn{1}{c}{<=} & \multicolumn{1}{c|}{>=} \\ \hline
1                       & 0.385                  & 0.652                  & 0.188                  & 0.839                  & 0.065                  & 0.947                   & 1.000                  & {\ul 0.001}            & 0.998                  & {\ul 0.003}            & 0.571                  & 0.429                   & 0.997                  & {\ul 0.005}            & \textbf{0.010}         & 0.993                  & 0.500                  & 0.539                   \\
2                       & 0.903                  & 0.116                  & 0.500                  & 0.539                  & \textbf{0.002}         & 0.999                   & 0.999                  & {\ul 0.002}            & 0.539                  & 0.500                  & \textbf{0.032}         & 0.976                   & 0.813                  & 0.216                  & 0.348                  & 0.688                  & 0.348                  & 0.688                   \\
3                       & 0.500                  & 0.539                  & 0.652                  & 0.385                  & 0.722                  & 0.313                   & \textbf{0.024}         & 0.981                  & \textbf{0.019}         & 0.986                  & \textbf{0.001}         & 1.000                   & 0.423                  & 0.615                  & 0.348                  & 0.688                  & 0.080                  & 0.935                   \\
4                       & \textbf{0.042}         & 0.968                  & 0.313                  & 0.722                  & 0.188                  & 0.839                   & 1.000                  & {\ul 0.001}            & 0.993                  & {\ul 0.010}            & 0.539                  & 0.500                   & 1.000                  & {\ul 0.001}            & 1.000                  & {\ul 0.001}            & 1.000                  & {\ul 0.001}             \\
5                       & 0.813                  & 0.216                  & 0.754                  & 0.278                  & 0.999                  & {\ul 0.002}             & 0.995                  & {\ul 0.007}            & 0.188                  & 0.839                  & \textbf{0.001}         & 1.000                   & 1.000                  & {\ul 0.001}            & 0.539                  & 0.500                  & 0.784                  & 0.246                   \\ \hline
6                       & 0.947                  & 0.065                  & \textbf{0.010}         & 0.993                  & \textbf{0.042}         & 0.968                   & 0.986                  & {\ul 0.019}            & \textbf{0.024}         & 0.981                  & 0.754                  & 0.278                   & 0.862                  & 0.161                  & \textbf{0.001}         & 1.000                  & \textbf{0.001}         & 1.000                   \\
7                       & 0.986                  & {\ul 0.019}            & \textbf{0.024}         & 0.981                  & \textbf{0.024}         & 0.981                   & 0.539                  & 0.500                  & 0.053                  & 0.958                  & \textbf{0.003}         & 0.998                   & 0.862                  & 0.161                  & \textbf{0.002}         & 0.999                  & \textbf{0.001}         & 1.000                   \\
8                       & \textbf{0.042}         & 0.968                  & \textbf{0.024}         & 0.981                  & \textbf{0.042}         & 0.968                   & \textbf{0.080}         & 0.935                  & \textbf{0.042}         & 0.968                  & 0.539                  & 0.500                   & 0.722                  & 0.313                  & 0.947                  & 0.065                  & 1.000                  & {\ul 0.001}             \\
9                       & 0.246                  & 0.784                  & 0.500                  & 0.539                  & \textbf{0.042}         & 0.968                   & \textbf{0.001}         & 1.000                  & \textbf{0.001}         & 1.000                  & \textbf{0.001}         & 1.000                   & 0.976                  & {\ul 0.032}            & 0.385                  & 0.652                  & 0.461                  & 0.577                   \\
10                      & 0.976                  & {\ul 0.032}            & 0.577                  & 0.461                  & 0.976                  & {\ul 0.032}             & \textbf{0.001}         & 1.000                  & \textbf{0.001}         & 1.000                  & \textbf{0.001}         & 1.000                   & 0.784                  & 0.246                  & 0.754                  & 0.278                  & 0.577                  & 0.461                   \\ \hline
11                      & 0.839                  & 0.188                  & 0.161                  & 0.862                  & \textbf{0.032}         & 0.976                   & 0.884                  & 0.138                  & 0.995                  & {\ul 0.007}            & 0.981                  & {\ul 0.024}             & 0.968                  & {\ul 0.042}            & \textbf{0.001}         & 1.000                  & \textbf{0.001}         & 1.000                   \\
12                      & 0.615                  & 0.423                  & 0.161                  & 0.862                  & 0.216                  & 0.813                   & 0.997                  & {\ul 0.005}            & 0.935                  & 0.080                  & 0.188                  & 0.839                   & 0.615                  & 0.423                  & 1.000                  & {\ul 0.001}            & 1.000                  & {\ul 0.001}             \\
13                      & 0.577                  & 0.461                  & 0.339                  & 0.661                  & 0.055                  & 0.945                   & \textbf{0.001}         & 1.000                  & \textbf{0.001}         & 1.000                  & \textbf{0.001}         & 1.000                   & 0.839                  & 0.188                  & 0.958                  & 0.053                  & 0.986                  & {\ul 0.019}             \\
14                      & 0.784                  & 0.246                  & 0.278                  & 0.754                  & 0.161                  & 0.862                   & \textbf{0.001}         & 1.000                  & \textbf{0.001}         & 1.000                  & \textbf{0.001}         & 1.000                   & 0.998                  & {\ul 0.003}            & 0.947                  & 0.065                  & 0.813                  & 0.216                   \\
15                      & 0.524                  & 0.476                  & 0.981                  & {\ul 0.024}            & 0.784                  & 0.246                   & \textbf{0.001}         & 1.000                  & \textbf{0.001}         & 1.000                  & \textbf{0.001}         & 1.000                   & 0.722                  & 0.313                  & 0.754                  & 0.278                  & 0.188                  & 0.839                   \\ \hline
\end{tabular}%
}

\end{table}


\newpage

Table \ref{tab:secondbudget} shows the results for the test function for the total budget of $10^6$. As expected the performance of all the models improved as we increased the total budget. Note that with the triangular noise, the relatively large MAPE measure is due to the small values (close to zero) for the true $\mathrm{CVaR}$. Moreover, since the additive noise term is bounded, the $\mathrm{CVaR}$ value can be accurately estimated using both POT and EMP. This results in a lower variance of $\mathrm{CVaR}$ and therefore all methods work relatively the same.

Table \ref{tab:thirdbudget} shows the results of the test function for the total budget of $10^7$. A relative improvement in performance is observed across all the different method-noise combinations in comparison to the last two tables, as we allocate a larger total budget. 


Table \ref{tab:pval} displays results from the Wilcoxon Signed Rank test, comparing our method with EMP-EMP. We focus on these two methods as ORD-KRG underperforms, and POT-EMP requires multiple replications at each design point.
To assess the relative performance we employ a one-sided hypothesis, rejecting the null hypothesis of a positive median difference at a $5\%$ confidence level. We also consider another one-sided hypothesis favoring EMP-EMP's performance over ours. Bolded values indicate the cases where the proposed approach outperformed EMP-EMP, while underlined values indicate the opposite.
In summary, there is a significant number of cases where each methodology is preferred. However, EVT-based estimation excels, particularly with larger $\alpha$ values in the case of Pareto noise (in 9 out of 15 cases and only one opposite result for $\alpha=0.995$), a crucial scenario due to the potential for significant losses from estimation errors.

\subsubsection{Stochastic network experiment}

\begin{figure}[!htb]
   \begin{minipage}{0.5\textwidth}
     \centering
     \includegraphics[width=.99\linewidth]{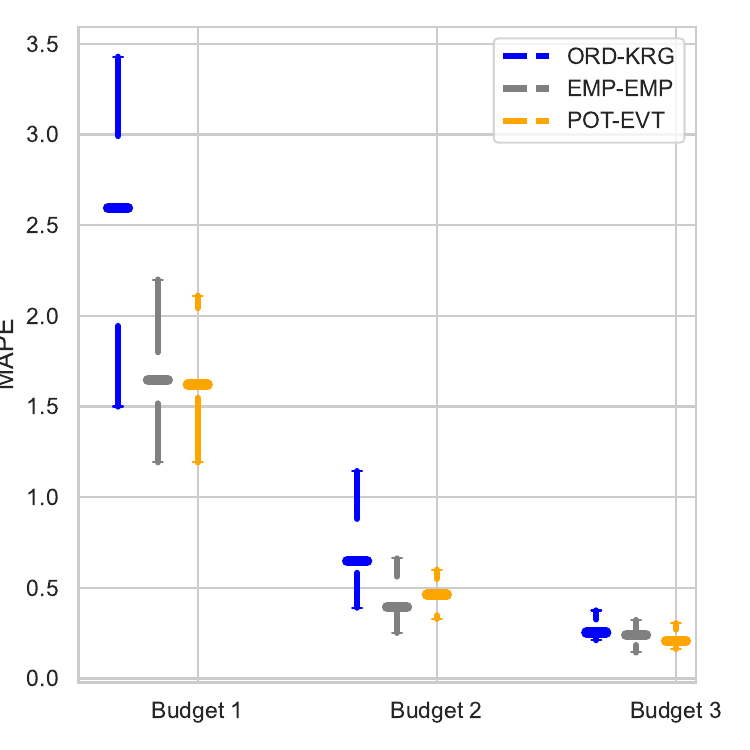}
     \caption{SAN results boxplots for ($\alpha = 0.95$) }\label{Fig:san95}
   \end{minipage}\hfill
   \begin{minipage}{0.5\textwidth}
     \centering
     \includegraphics[width=.99\linewidth]{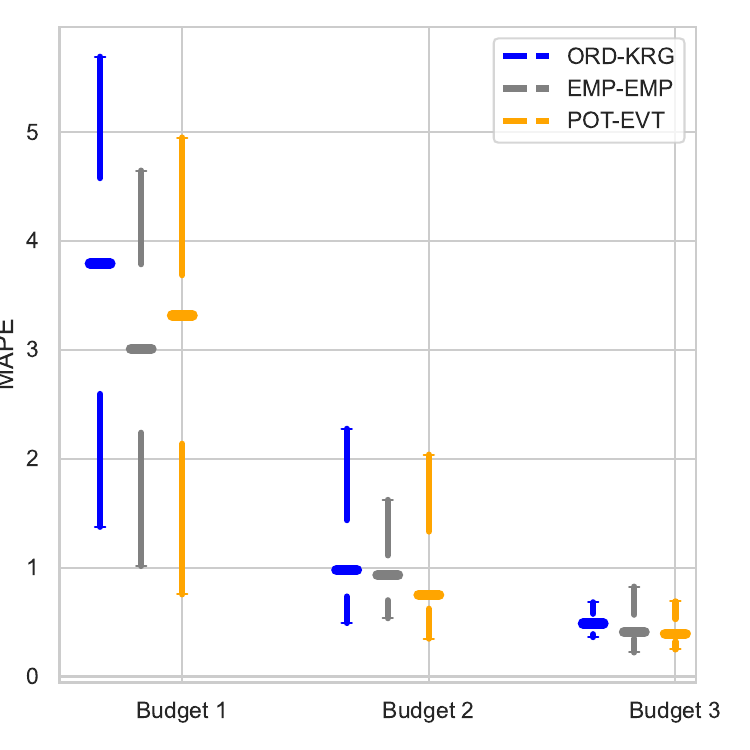}
     \caption{SAN results boxplots for ($\alpha = 0.99$)}\label{Fig:san99}
   \end{minipage}\hfill
\end{figure}

\begin{figure}[!htb]
     \centering
     \includegraphics[width=.5\linewidth]{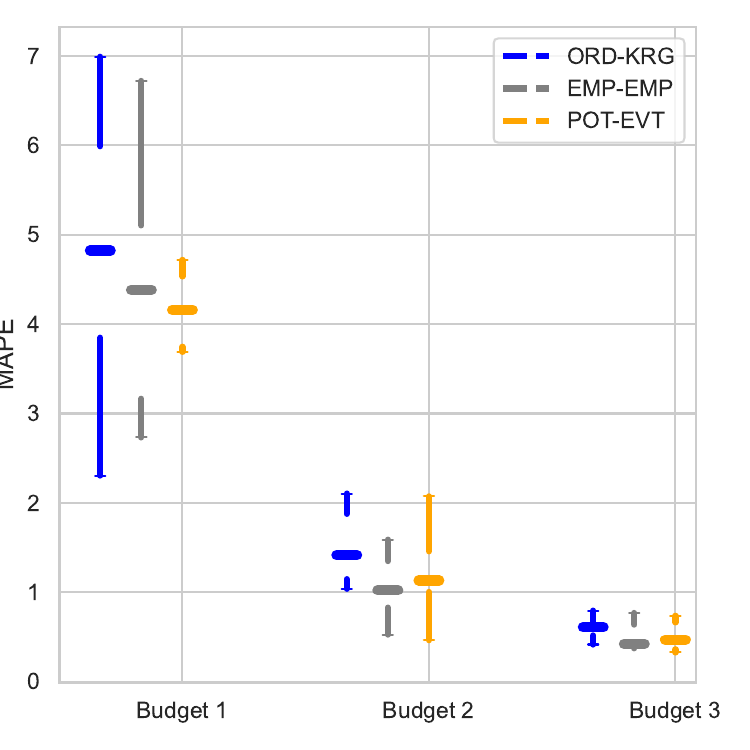}
     \caption{SAN results boxplots for ($\alpha = 0.995$)}\label{Fig:san995}
\end{figure}

Table \ref{tab:SAN_results} shows the results for the SAN problem for three different $\alpha$ levels, 0.95, 0.99, and 0.995. 193 (resulting in a total of 200, including 7 design points) equally spaced points are chosen from the domain space to evaluate the performance of the surrogate model. The
true value of $\mathrm{CVaR}$ at each of these design points is obtained
numerically by inverting the CDF of the random project time. The obtained MAPE measures for each different $\alpha$ level are provided as a box plot (Figure \ref{Fig:san95}-- \ref{Fig:san995})  to graphically depict groups of MAPE measures through their quartiles.

\begin{table}[]
\centering
\caption{Median of 10 MAPE measures for predicting $\mathrm{CVaR}$ of SAN over the test set }
\label{tab:SAN_results}
\resizebox{.5\textwidth}{!}{%
\begin{tabular}{|c|c|lll|}
\hline
 &  & \multicolumn{3}{c|}{alpha} \\ \cline{3-5} 
\multirow{-2}{*}{Method} & \multirow{-2}{*}{Budget} & \multicolumn{1}{c}{0.95} & \multicolumn{1}{c}{0.99} & \multicolumn{1}{c|}{0.995} \\ \hline
ORD-KRG &  & 2.59 & 3.79 & 4.82 \\
EMP-EMP &  & 1.62 & 3.32 & 4.16 \\
POT-EVT & \multirow{-3}{*}{\rotatebox[origin=c]{90}{1000 }} & 1.65 & 3.01 & 4.38 \\ \hline
ORD-KRG &  & \cellcolor[HTML]{FFFFFF}0.65 & \cellcolor[HTML]{FFFFFF}0.98 & \cellcolor[HTML]{FFFFFF}1.41 \\
EMP-EMP &  & \cellcolor[HTML]{FFFFFF}0.46 & \cellcolor[HTML]{FFFFFF}0.75 & \cellcolor[HTML]{FFFFFF}1.13 \\
POT-EVT & \multirow{-3}{*}{\rotatebox[origin=c]{90}{10,000 }} & \cellcolor[HTML]{FFFFFF}0.39 & \cellcolor[HTML]{FFFFFF}0.93 & \cellcolor[HTML]{FFFFFF}1.02 \\ \hline
ORD-KRG &  & \cellcolor[HTML]{FFFFFF}0.25 & \cellcolor[HTML]{FFFFFF}0.49 & \cellcolor[HTML]{FFFFFF}0.61 \\
EMP-EMP &  & \cellcolor[HTML]{FFFFFF}0.21 & \cellcolor[HTML]{FFFFFF}0.39 & \cellcolor[HTML]{FFFFFF}0.47 \\
POT-EVT & \multirow{-3}{*}{\rotatebox[origin=c]{90}{100,000 }} & \cellcolor[HTML]{FFFFFF}0.24 & \cellcolor[HTML]{FFFFFF}0.41 & \cellcolor[HTML]{FFFFFF}0.42 \\ \hline
\end{tabular}%
}

\end{table}



\begin{table}[!h]
\centering
\caption{P-value results for the one-sided Wilcoxon Signed Rank Test over the MAPE of the POT-EVT and EMP-EMP SK models. Bolded values support our claim that our model has a lower error, while the underlined values show that EMP-EMP method achieves a lower error.}
\label{tab:sanpval}
\resizebox{.5\textwidth}{!}{%
\begin{tabular}{|r|llllll|}
\hline
\multirow{3}{*}{Budget} & \multicolumn{6}{c|}{$\alpha$}                                                                                                                                   \\ \cline{2-7} 
                        & \multicolumn{2}{c|}{0.95}                           & \multicolumn{2}{c|}{0.99}                           & \multicolumn{2}{c|}{0.995}                       \\ \cline{2-7} 
                        & \multicolumn{1}{c}{<=} & \multicolumn{1}{c|}{>=}    & \multicolumn{1}{c}{<=} & \multicolumn{1}{c|}{>=}    & \multicolumn{1}{c}{<=} & \multicolumn{1}{c|}{>=} \\ \hline
1000                    & 0.278                  & \multicolumn{1}{l|}{0.754} & 0.539                  & \multicolumn{1}{l|}{0.500} & 0.138                  & 0.884                   \\
10000                   & 0.839                  & \multicolumn{1}{l|}{0.188} & 0.385                  & \multicolumn{1}{l|}{0.652} & 0.754                  & 0.278                   \\
100000                  & 0.313                  & \multicolumn{1}{l|}{0.722} & \textbf{0.032}         & \multicolumn{1}{l|}{0.976} & 0.161                  & 0.862                   \\ \hline
\end{tabular}%
}

\end{table}

Table \ref{tab:sanpval} shows the p-values resulting from the one-sided and two-sided Wilcoxon Signed Rank test over the difference of MAPE measure between the POT-EVT method and EMP-EMP method. As shown the two methodologies perform mostly the same because the intrinsic variability of  $\mathrm{CVaR}$ estimations is relatively small. \blue{ Also, it can be observed in Table \ref{tab:SAN_results} that as the number of observations increases, the performance of the ordinary kriging model (with no intrinsic variance) becomes relatively similar to the other methodologies due to the fact that the variance of the $\mathrm{CVaR}$ estimations drops. 
}

\section{Conclusions and Future Research }


In this work, we investigated the potential of metamodeling combined with extreme value theory to address risk management in simulation models. We used Stochastic kriging to globally characterize tail behavior, and we developed a framework with efficient methods for estimating intrinsic variance without multiple replications.
To test and evaluate our proposed method, we used two numerical examples with different budget allocations and noise levels. We compared our proposed methodology with other methods in the literature and generally observed that using POT estimation of $\mathrm{CVaR}$ and its approximate asymptotic variance for stochastic spatial functions with a  fat-tailed additive noise at high $\alpha$ levels outperforms the existing empirical approaches. 

Although we perform numerical tests under different scenarios with the two benchmark examples, it nevertheless is necessary to validate our methodology with large and complex simulation models of real-world problems. We expect that numerical issues 
would arise in fitting high-dimensional kriging models. \blue{Moreover, we consider stochastic kriging without the trend term. Naturally, in some applications, using a form of polynomial trend term can result in better predictions. While we expect that a similar methodology can be employed in such cases, this is not explicitly tested here.} In this work we use MLE to estimate the $\mathrm{GPD}$ parameters, however, other estimation methods such as the method of Probability-Weighted Moments and the method of Moments can be useful methods to look into as they are shown to be more reliable with smaller sample size \citep{hosking1987parameter}. 

Lastly,  POT is not the only existing approach to evaluating extreme behavior of stochastic systems. Block maxima approach is another popular approach within the general extreme value theory framework. Both block maxima and POT have advantages and disadvantages and are widely used \citep[see, for example, ][specifically in the context  of risk measurement]{tsay2005analysis}. A thorough investigation of the potential to employ block maxima for CVaR or other risk measure estimation for metamodelling application is beyond the  scope of this study and could be a topic for a future study. 


\section*{Supplementary materials}
All code required for implementing the proposed method and that was used for the numerical experiments is available at  \url{https://github.com/arminkhayyer/EVT_CVaR_SK/tree/master}.

\section*{Disclosure statement}

All authors certify that they have no affiliations with or involvement in any organization or entity with any financial interest or non-financial interest in the subject matter or materials discussed in this manuscript.

\section*{Funding}

Part of this work was funded by the project ``SBIR Phase III – Analytical Framework and Modeling to Support Wargaming Logistics'' by Frontier Technology, Inc.





\bibliographystyle{natbib}  
\bibliography{references}  

\section{Appendix}
\begin{figure}[!htb]

     \centering
     \includegraphics[width=.99\linewidth]{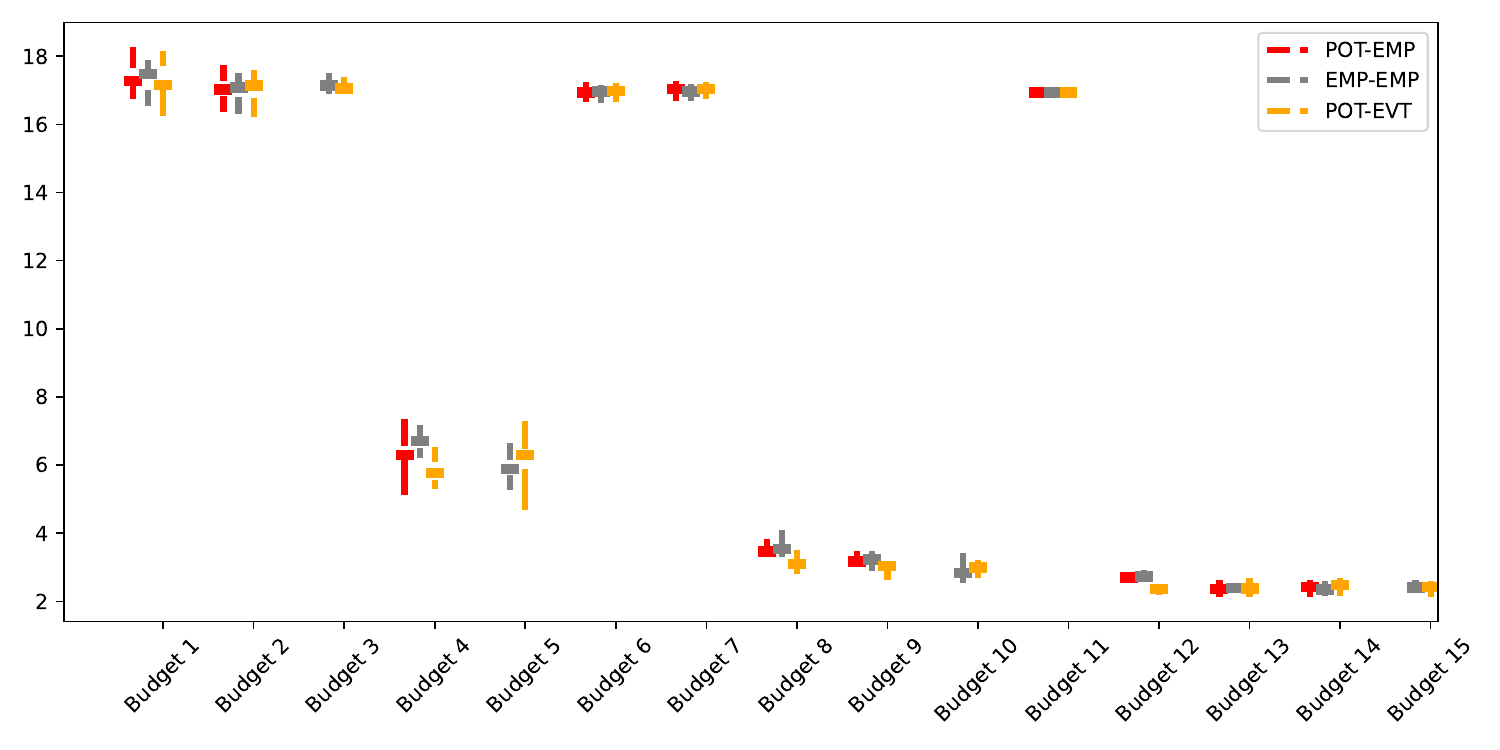}
     \caption{boxplots for the normal noise $\alpha = 0.95 $ }\label{Fig:normal95}
\end{figure}
\begin{figure}[!htb]
     \centering
     \includegraphics[width=.99\linewidth]{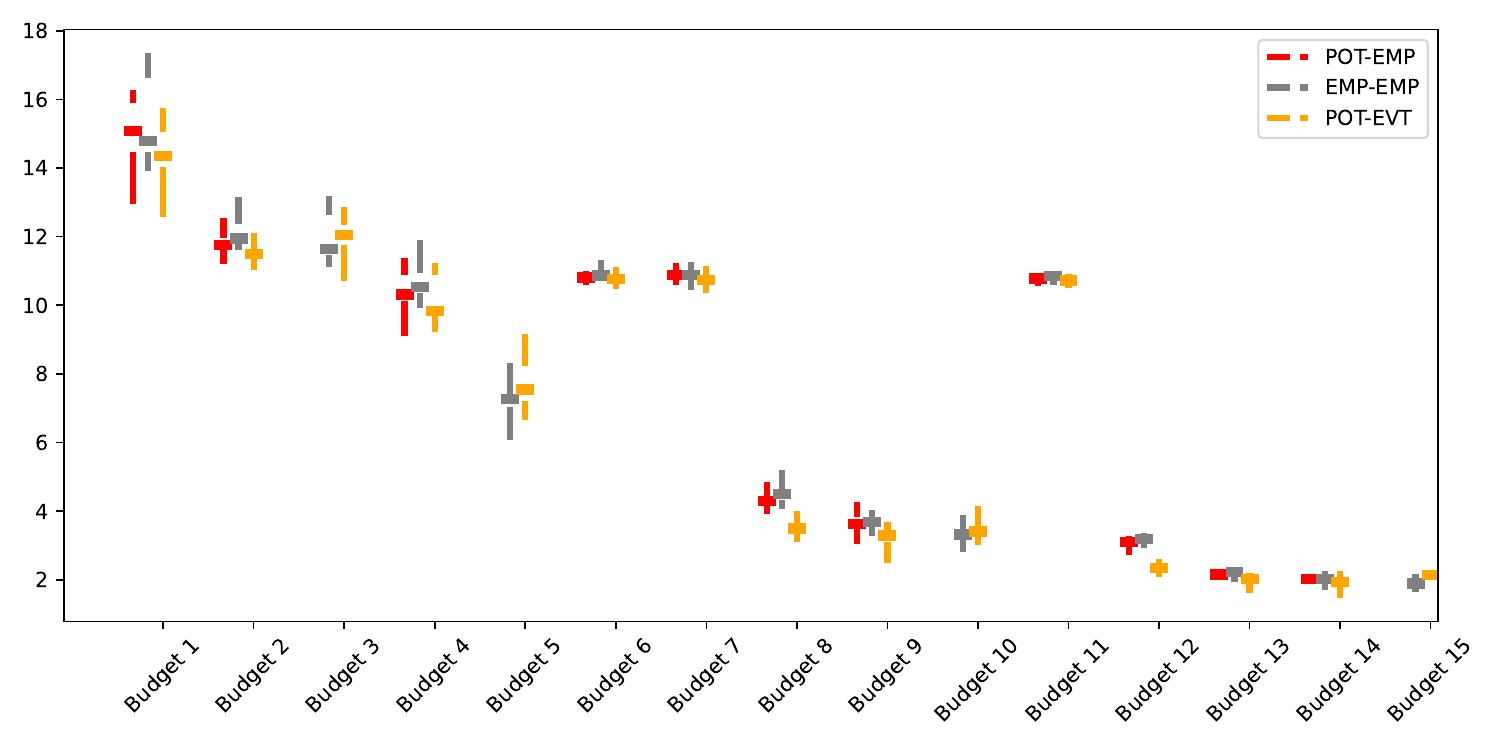}
     \caption{boxplots for the normal noise $\alpha = 0.995 $}\label{Fig:normal995}

\end{figure}

\begin{figure}[!htb]
     \centering
     \includegraphics[width=.99\linewidth]{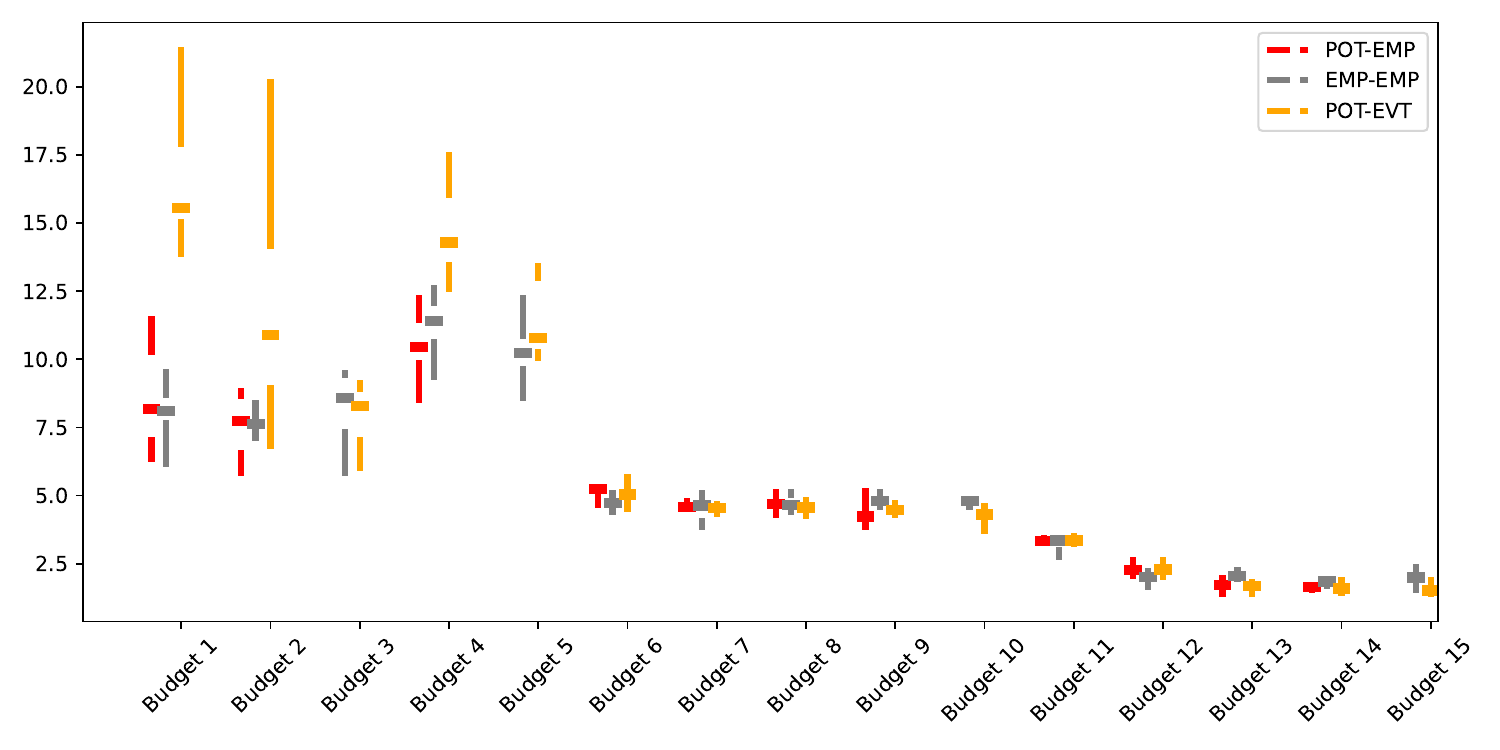}
     \caption{boxplots for the pareto noise $\alpha = 0.95 $ }\label{Fig:pareto95}
\end{figure}
\begin{figure}[!htb]
     \centering
     \includegraphics[width=.99\linewidth]{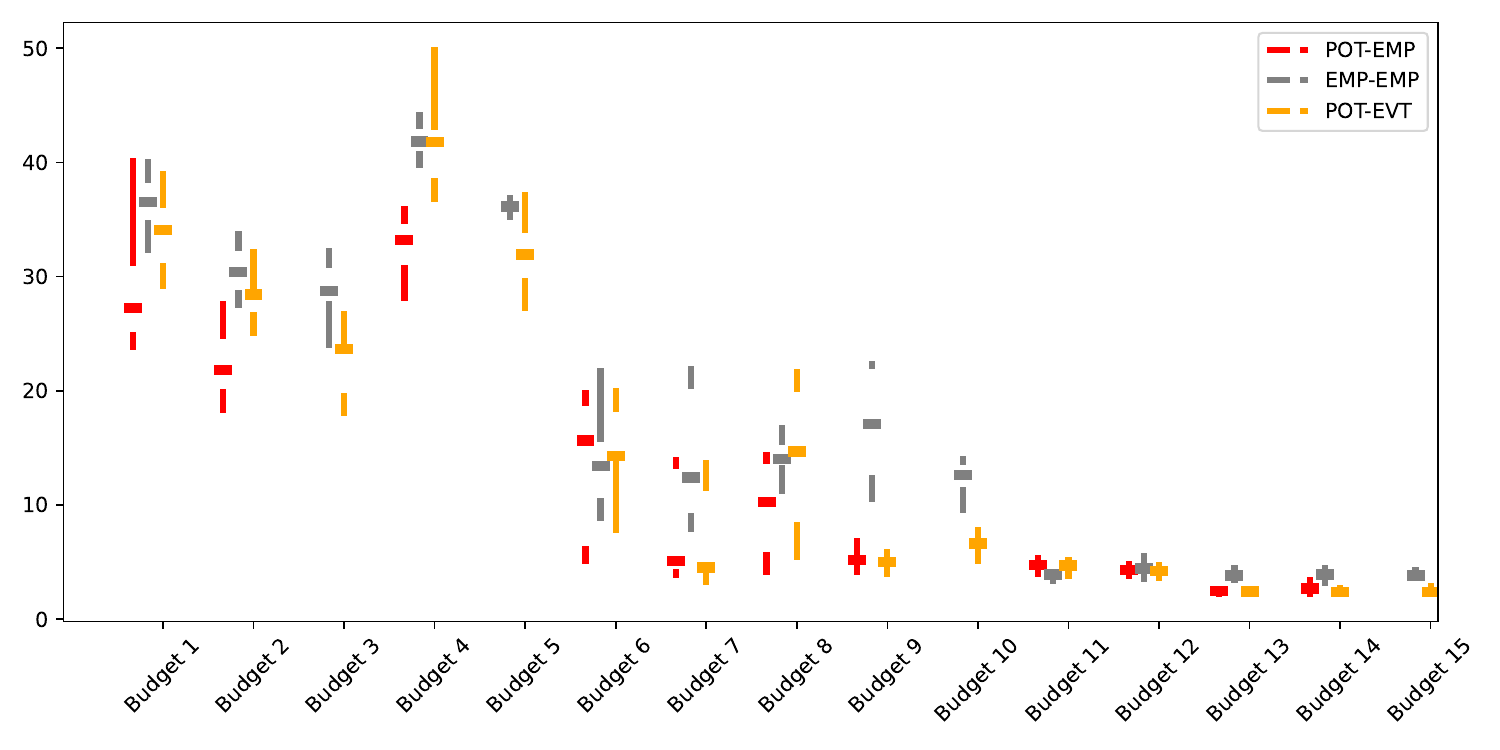}
     \caption{boxplots for the pareto noise $\alpha = 0.995 $}\label{Fig:pareto995}
\end{figure}

\begin{figure}[!htb]
     \centering
     \includegraphics[width=.99\linewidth]{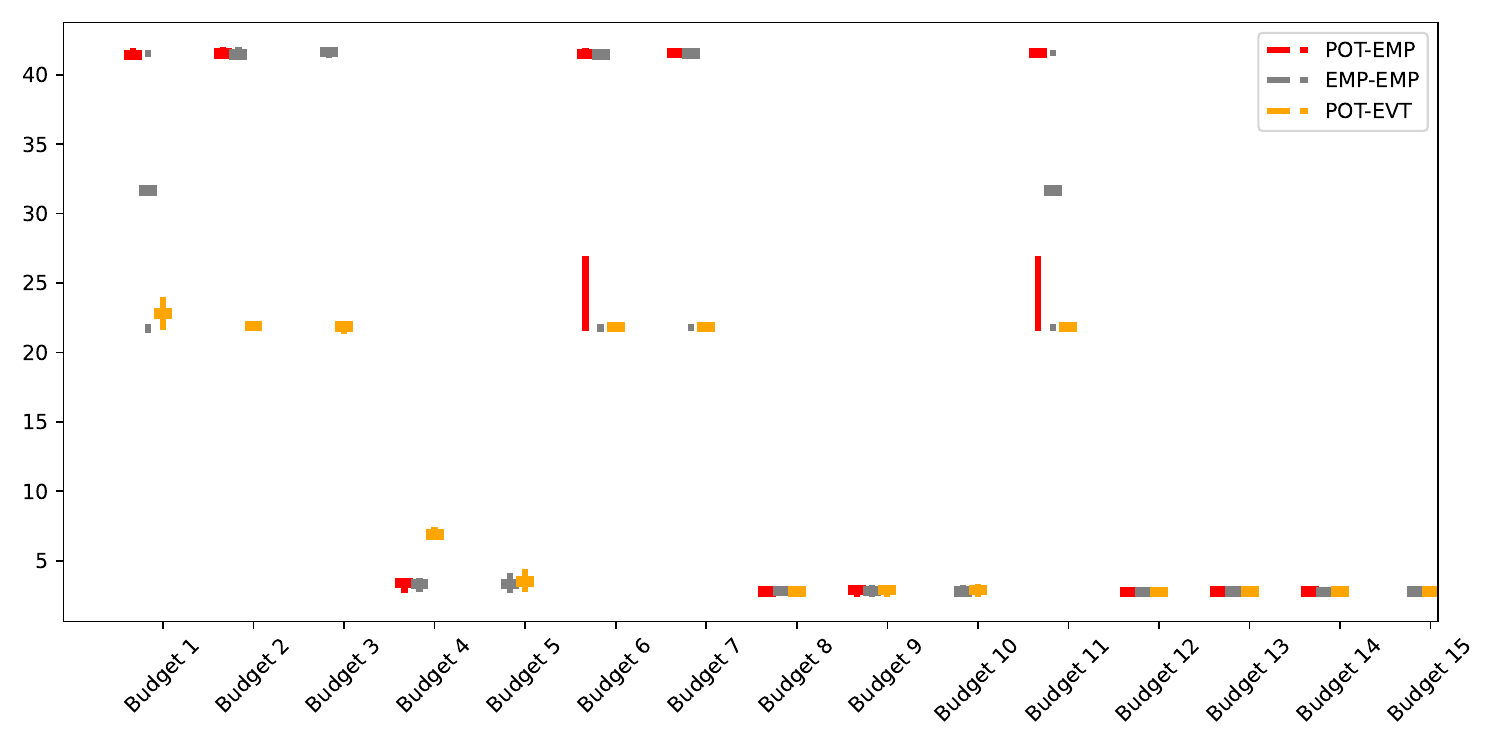}
     \caption{boxplots for the Triangular noise $\alpha = 0.95 $ }\label{Fig:traing95}
\end{figure}
\begin{figure}[!htb]
     \centering
     \includegraphics[width=.99\linewidth]{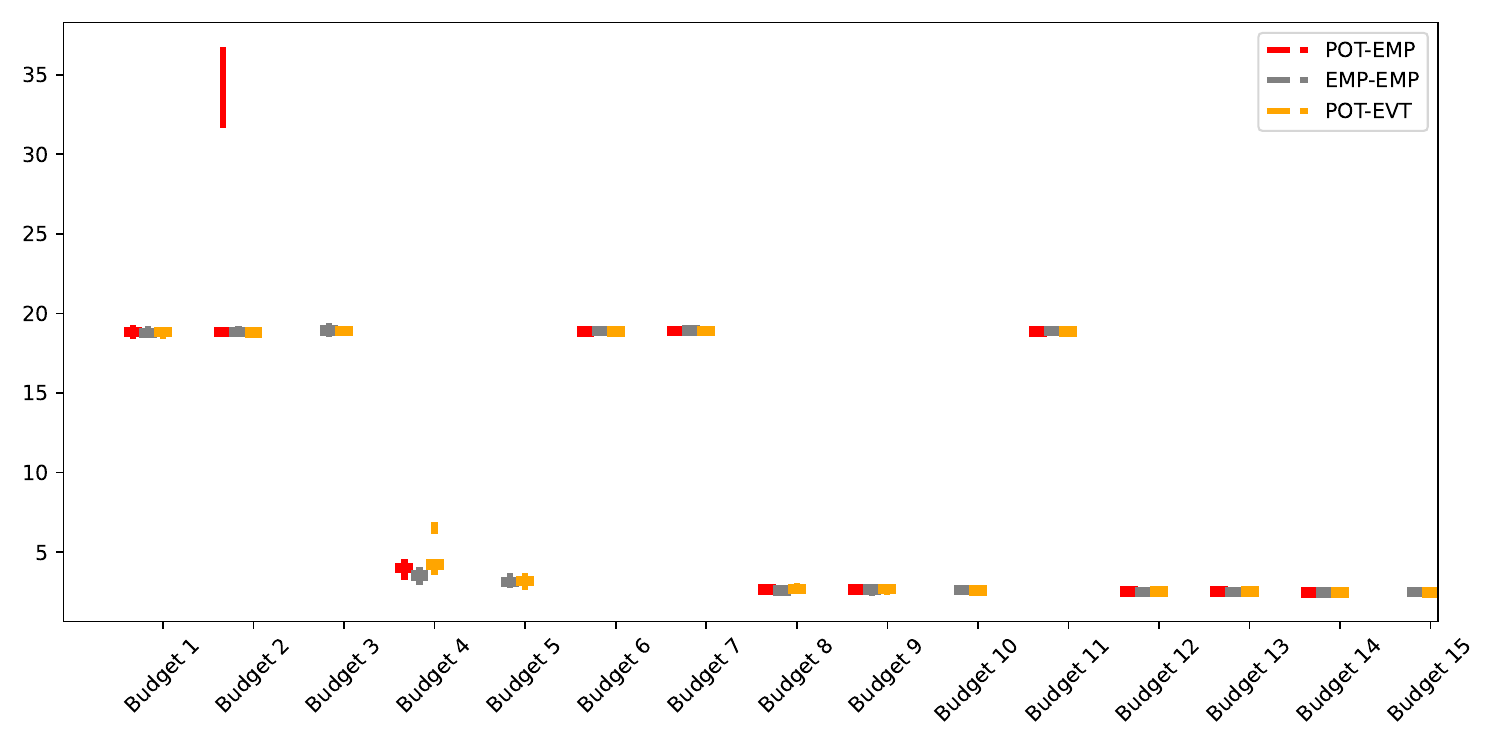}
     \caption{boxplots for the Triangular noise $\alpha = 0.995 $}\label{Fig:traing995}
\end{figure}

\end{document}